\documentclass[acmsmall,screen]{acmart}

\AtBeginDocument{%
  \providecommand\BibTeX{{%
    \normalfont B\kern-0.5em{\scshape i\kern-0.25em b}\kern-0.8em\TeX}}}

\renewcommand\footnotetextcopyrightpermission[1]{} 

\setcopyright{acmlicensed}

 \usepackage{graphicx}
 \usepackage{colortbl}
 \usepackage{soul} 
 \usepackage{color, xcolor}
\usepackage{enumitem}
\usepackage{multirow}

\usepackage{booktabs,arydshln}
\usepackage{tcolorbox}
\usepackage{makecell}
\usepackage{array}
\usepackage{geometry}
\usepackage{longtable}
\usepackage{lscape}
\newcommand{\name}{{\textit{CausaDisco}}}
\newcommand{\rr}[1]{{\color{black} #1}}
\newcommand{\yf}[1]{{\color{black} #1}}

\begin{document}

\title[Epistemologically-Informed LLM for Self-Learning]
{Enhanced Self-Learning with Epistemologically-Informed LLM Dialogue}



\author{Yi-Fan Cao}
\orcid{0000-0002-5892-5052}
\affiliation{%
  \institution{Hong Kong University of Science and Technology}
  \city{Hong Kong}
  \country{China}}
\email{ycaoaw@connect.ust.hk}

\author{Kento Shigyo}
\orcid{0000-0002-5095-7500}
\affiliation{%
  \institution{Hong Kong University of Science and Technology}
  \city{Hong Kong}
  \country{China}}
\email{kshigyo@connect.ust.hk}

\author{Yitong Gu}
\orcid{0009-0001-2890-5448}
\affiliation{%
  \institution{Hong Kong Baptist University}
  \city{Hong Kong}
  \country{China}}
\email{yitonggu@life.hkbu.edu.hk}

\author{Xiyuan Wang}
\orcid{0009-0008-1839-2010}
\affiliation{%
  \institution{ShanghaiTech University}
  \city{Shanghai}
  \country{China}}
\email{wangxy7@shanghaitech.edu.cn}

\author{Weijia Liu}
\orcid{0009-0001-6332-5705}
\affiliation{%
  \institution{Hong Kong University of Science and Technology (Guangzhou)}
  \city{Guangzhou}
  \country{China}}
\email{wliu383@connect.hkust-gz.edu.cn}

\author{Yang Wang}
\affiliation{%
  \institution{The University of Hong Kong}
  \city{Hong Kong}
  \country{China}}
\email{yang.wang@hku.hk}
\orcid{0000-0002-8903-2388}

\author{David Gotz}
\orcid{0000-0002-6424-7374}
\affiliation{%
  \institution{University of North Carolina at Chapel Hill}
  \city{Chapel Hill}
  \state{North Carolina}
  \country{USA}}
\email{gotz@unc.edu}

\author{Zhilan Zhou}
\authornote{Zhilan Zhou is the corresponding author of this research.}
\orcid{0000-0003-1236-1287}
\affiliation{%
  \institution{University of North Carolina at Chapel Hill}
  \city{Chapel Hill}
  \state{North Carolina}
  \country{USA}}
\email{zzl@cs.unc.edu}

\author{Huamin Qu}
\affiliation{%
  \institution{Hong Kong University of Science and Technology}
  \city{Hong Kong}
  \country{China}}
\email{huamin@ust.hk}
\orcid{0000-0002-3344-9694}

\renewcommand{\shortauthors}{Cao et al.}
\renewcommand\arraystretch{1.2}

\begin{abstract}

Large Language Models (LLMs) have advanced self-learning tools, enabling more personalized interactions. However, learners struggle to engage in meaningful dialogue and process complex information. To alleviate this, we incorporate epistemological frameworks within an LLM-based approach to self-learning, reducing the cognitive load on learners and fostering deeper engagement and holistic understanding.
\rr{Through a formative study (N=26), we identified epistemological differences in self-learner interaction patterns. Building upon these findings, we present \textit{CausaDisco}, a dialogue-based interactive system that integrates Aristotle's \textit{Four Causes} framework into LLM prompts to enhance cognitive support for self-learning.} This approach guides learners' self-learning journeys by automatically generating coherent and contextually appropriate follow-up questions. A controlled study (N=36) demonstrated that, compared to baseline, \textit{CausaDisco} fostered more engaging interactions, inspired sophisticated exploration, and facilitated multifaceted perspectives. This research contributes to HCI by expanding the understanding of LLMs as educational agents and providing design implications for this emerging class of tools.

\end{abstract}



\begin{teaserfigure}
  \includegraphics[width=\textwidth]{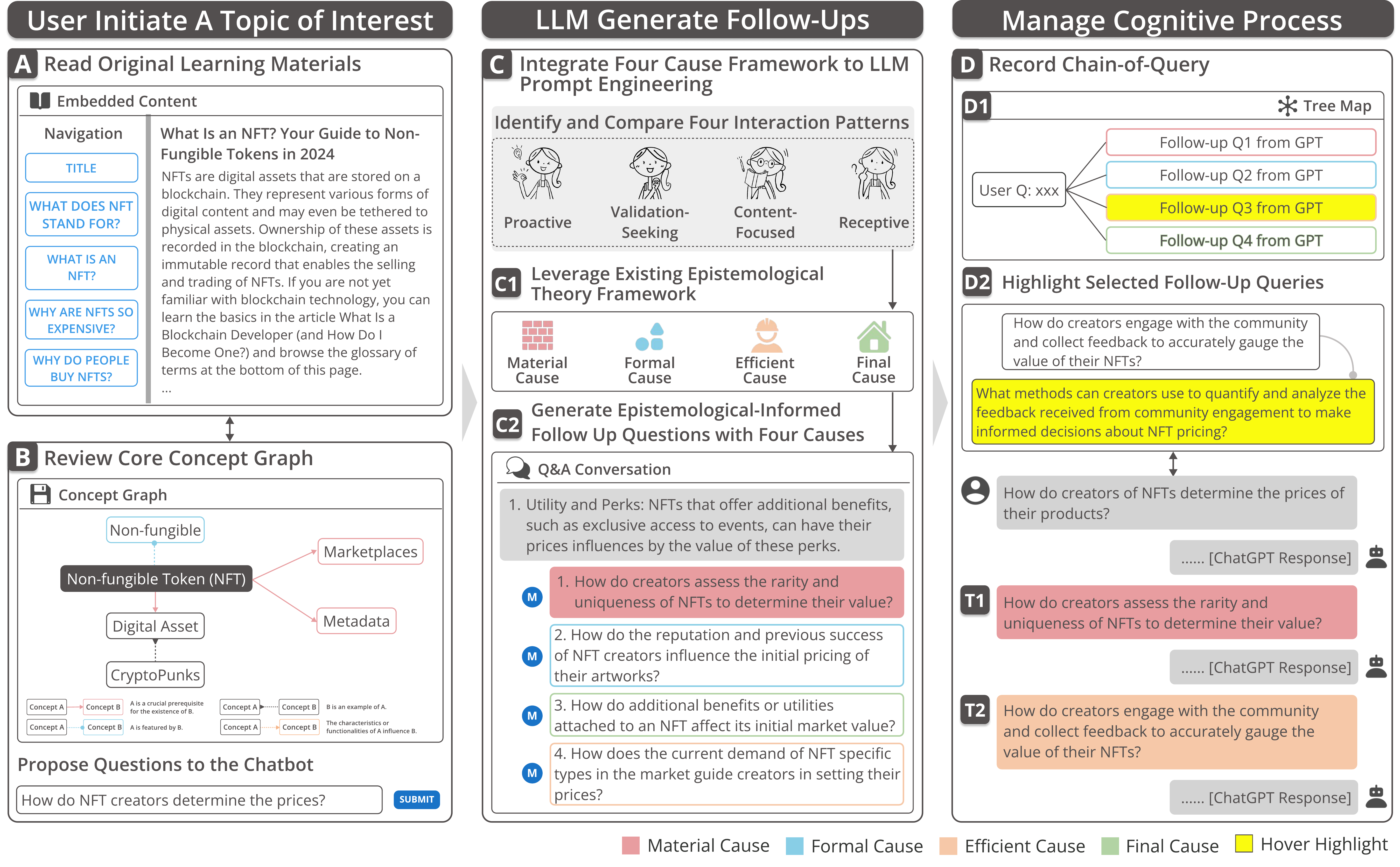}
  \caption{\textit{CausaDisco} provides users with original learning materials (A) and a core concept graph (B) to facilitate self-learning. 
  Once a user initiates a dialogue, the LLM chatbot offers preliminary answers and then automatically generates four epistemologically-informed follow-up questions (C) to encourage deeper exploration. Concurrently, \textit{CausaDisco} creates a query tree map (D) to assist users in managing their conversation records.
  When users are satisfied with the results of multi-turn dialogues (T1-3), they can proceed to explore and learn about new topics.}
  \Description{This figure presents the interface and interaction design of CausaDisco, which enhances self-learning by providing original materials and a core concept graph for guidance. When users start a dialogue, the LLM chatbot offers initial answers and generates four follow-up questions to promote deeper understanding. Simultaneously, a query tree map is created to help users manage conversation records. Once users are satisfied with multi-turn dialogues, they can explore new topics.}
  \label{fig:teaser}
\end{teaserfigure}


\maketitle

\section{Introduction}

The rise of Large Language Models (LLMs) has revolutionized education, introducing a new generation of dialogue-driven learning tools \cite{luo2022critical, chen2023artificial, zhang2023visar}.
These tools have greatly improved the efficiency of independent learning, supporting a wide range of tasks, from spoken language practice \cite{huang2022chatbots} and text editing \cite{kim2023using} to literature searches \cite{zheng2024charting, zheng2024disciplink} and information summarization \cite{reddy2023automatic}.
With the adoption of prompt engineering techniques, the potential of LLM-based educational tools to facilitate deeper learning and knowledge construction has been further explored \cite{suh2023sensecape}.
These dialogue-driven tools can help streamline the cognitive processes involved in sensemaking, enabling learners to acquire, comprehend, integrate, and apply new knowledge more effectively~\cite{abdelghani2024gpt, gero2024supporting}.

Although LLM-based educational tools promote a flexible and intuitive ``search-as-learning'' sensemaking process \cite{liu2024selenite}, many still require substantial effort from learners to engage in meaningful dialogue and process complex information.
This is particularly challenging when using less customized LLM tools, as learners often spend considerable time crafting precise questions to elicit responses aligned with their specific needs.
Such exploratory and iterative dialogue-driven interactions can be especially demanding for less proactive learners accustomed to more passive input-driven sensemaking.
Consequently, even with LLM-based tools, many learners still encounter substantial, yet often hidden, cognitive load.

Extensive studies have focused on enhancing the contextuality \cite{8808076, shihab2023revisiting, chen2023gap} and workflow \cite{suh2023sensecape, gao2024fine, shankar2024validates} of LLM-based educational tools to support learners' sensemaking processes.
However, \rr{existing work often centers} on domain-specific applications \cite{sheng2023knowledge, chen2024stugptviz} or caters to early adopters already proficient in strategically engaging with LLM chatbots \cite{zheng2024disciplink}.
\rr{This focus overlooks a fundamental aspect of human-AI interaction: sensemaking is a highly individualized cognitive process.
This process is shaped by one's \textit{epistemological schema}, which represents structured beliefs about the \textit{nature}, \textit{source}, and \textit{justification} of knowledge~\cite{hofer1997development, brownlee2002core, braaten2010personal, baker2020epistemic}.}
Consequently, a critical gap exists in understanding how diverse mental models, reflected in different LLM interaction behaviors influence the design of effective, usable, and inclusive educational tools.

To bridge this knowledge gap, we conducted a formative study with 26 participants.
\rr{Data collection was triangulated~\cite{carter2014use} through dialogue records captured by a probe system, online surveys, and semi-structured interviews~\footnote{Please refer to the supplementary materials for details of our survey and semi-structured interview protocols.}. Through a thematic analysis of the collected data, we derived a query taxonomy that revealed distinct interaction behaviors, while also examining participants' underlying epistemological schemas. The synthesis of these analyses identified four \textit{interaction patterns}: \textit{proactive}, \textit{validation-seeking}, \textit{content-focused}, and \textit{receptive}.
We revealed key differences by analyzing these patterns within the epistemological framework of knowledge nature, source, and justification, particularly between proactive and receptive patterns.
The proactive pattern was characterized by active knowledge construction, evidenced by participants formulating effective questions, requesting elaborations, and synthesizing information. Conversely, the receptive pattern was characterized by fewer questions, less exploration, and a preference for receiving direct answers from LLMs rather than actively building understanding. Furthermore, the proactive pattern was associated with high-performing participants, while the receptive pattern was often observed in those who struggled in the learning process.}

\rr{The observed epistemological differences and learning outcomes across different patterns highlight the need for tools to accommodate diverse learner needs. We identified three primary challenges in designing inclusive learner-LLM interactions}: \textit{limited interactivity}, \textit{inefficient information verification}, and \textit{confirmation bias}.
Addressing these challenges requires educational tools that support: (1) lower cognitive load, (2) sustained engagement, (3) multidimensional thinking, (4) metacognition, and (5) validation assistance.
Guided by these requirements, we developed \textit{CausaDisco}, a proof-of-concept system designed to support more effective learner-LLM interactions.
\textit{CausaDisco} comprises four main views (see Fig. \ref{fig:teaser}): an \textit{Embedded Content View} providing access to original learning materials; a \textit{Concept Graph} visualizing relationships between core concepts; a \textit{Q\&A Conversation View} prompting epistemologically-informed follow-up questions; and a \textit{Tree Map View} displaying the underlying query logic. 

Specifically, in the \textit{Q\&A Conversation View}, we incorporate epistemological frameworks with LLM prompt engineering to enhance self-learners sensemaking.
\rr{This approach is inspired by pedagogical research demonstrating the benefits of structured, epistemologically-informed frameworks for critical thinking, analytical depth, and knowledge acquisition in self-learning contexts} \cite{peters2000does, braaten2010personal, krasmann2020logic}.
Drawing on these insights, as well as \rr{the interaction patterns distilled from} our formative study, we identified the need for a systematic approach to generating follow-up questions that support comprehensive sensemaking.
We selected Aristotle's \textit{Four Causes} \cite{hocutt1974aristotle, falcon2006aristotle}--a classical epistemological framework for knowledge construction--due to its remarkable alignment with our empirically derived query taxonomy.
This framework prompts users to consider the \textit{Material}, \textit{Formal}, \textit{Efficient}, and \textit{Final} causes of concepts, effectively mapping to the diverse types of questions learners naturally ask.

To evaluate the efficacy of \textit{CausaDisco}, we conducted a controlled user study with 36 participants employing a between-subjects design. 
We assessed both objective and subjective measures, including quiz scores and Likert-scale ratings to assess engagement, efficiency, comprehensiveness, sophistication, and usability of the system.
Our results indicate that \textit{CausaDisco} significantly facilitates users' sensemaking process, fostering a more engaging learning experience compared to the baseline condition.
In summary, our core contributions are as follows:

\begin{itemize}

\item 
A taxonomy of interaction patterns in self-learning with LLMs, identifying key epistemological differences and outlining design requirements for supporting user sensemaking.

\item
\textit{CausaDisco}, an interactive proof-of-concept system for self-learning. The system leverages Aristotle's \textit{Four Causes} theoretical framework for LLM prompt engineering to guide user sensemaking and includes features for knowledge synthesis and learning journey tracking.

\item
Empirically-grounded design implications for dialogue-driven LLM educational tools, derived from a within-subjects evaluation of \textit{CausaDisco}'s efficacy in enhancing interactivity and sensemaking.
\end{itemize}

\section{Related Work}

\rr{Given the interdisciplinary nature of this research, our literature review integrates technical, epistemological, and pedagogical perspectives. It is structured as follows:}
1) \textit{LLM-based educational agents}; 2) \textit{Integrating epistemology for LLM sensemaking}; and 3) \textit{Toward effective learner-LLM interactions}.

\subsection{LLM-Based Educational Agents}

\rr{LLM-based educational agents advance technology-enhanced learning (TEL) by creating personalized learning experiences~\cite{gan2023large, lin2023exploring, huber2024leveraging}.}
These agents improve intelligent tutoring systems (ITS)  through real-time assistance~\cite{li2024bringing}, customized feedback~\cite{fu2024examining}, and adaptive learning paths~\cite{lee2024few}. Such features empower learners to regulate their cognitive processes and support various tasks, including self-reflection~\cite{Li_Liang_Le_Lc_Luo_2023}, goal-setting~\cite{Deng_Lei_Huang_Chua_2023}, and problem-solving~\cite{kumar2023impact}. 
\rr{This makes LLM-based educational agents particularly valuable in self-learning contexts, where individualized and adaptive support is critical.}

\rr{Self-directed learning is a process involving continuous cognitive and behavioral adjustment aimed at understanding specific knowledge domains~\cite{demirbag2021modeling, huang2023students}. Learning outcomes can vary depending on individual strategies, and pedagogical research emphasizes the importance of active engagement and self-regulation in this process~\cite{wang2015investigating, wekerle2022using}. Recent HCI studies have explored how LLM-based educational agents can enhance the effectiveness of self-learning by promoting autonomy, particularly from students' perspectives~\cite{limna2023use, li2024large}.
For instance, Goslen \textit{et al.}\cite{goslen2024llm} proposed an LLM-based system that automatically generates student planning strategies based on historical interaction data, enabling dynamic support in game-based learning environments.
Through fine-tuning for domain-specific instruction, these agents have demonstrated success in areas such as language learning~\cite{liu2024scaffolding}, technology literacy~\cite{gao2024fine, hartley2024artificial}, and research ideation~\cite{liu2024personaflow}. These advances enable self-learners to effectively manage their cognitive processes and improve practical skills with reduced dependence on human guidance.

Despite these advances, LLM-based educational agents have yet to achieve the dynamic reciprocal interaction where learners and systems mutually shape each other's development, as envisioned by pedagogical scholars~\cite{schulte2018framework}.
This limitation stems primarily from the agents' inability to guide users in crafting effective prompts ~\cite{shah2022situating, fiannaca2023programming, zamfirescu2023johnny}, which increases cognitive load and hinders knowledge acquisition.
Although features such as automatic question generation~\cite{zheng2024disciplink, hu2024designing} and dialogue history management~\cite{safranek2023role} improve interaction intuitiveness, a lack of systematic guidance limits the ability of self-learners---especially those unfamiliar with AI or interactive learning---to fully harness these agents' potential.} Our study addresses this gap by investigating how different mental models shape self-learners interaction patterns with LLM-based educational agents. These insights will inform the design of more effective, accessible, and inclusive AI-enhanced educational systems, ultimately enhancing learner engagement for a broader audience.

\rr{\subsection{Integrating Epistemology for LLM Sensemaking}}

The advent of LLMs has spurred their integration into tools augmenting human sensemaking during self-learning.
\rr{Sensemaking is a vital cognitive process that influences how learners process information, ultimately impacting their learning outcomes and efficiency~\cite{marchionini2019search, rong2023understanding}.}
This cyclical process encompasses several cognitive stages: \textit{acquiring}, \textit{comprehending}, \textit{integrating}, and \textit{applying} new knowledge~\cite{pirolli2005sensemaking}. Recognizing LLMs’ potential to transform this process, the HCI research community has explored various approaches to enhance the different cognitive stages~\cite{lee2024hints, gero2024supporting, ma2024beyond}.
For instance, Zheng \textit{et al.}~\cite{zheng2024disciplink} enhanced knowledge search and acquisition through an interdisciplinary literature search system that facilitates cross-domain synthesis. Building on this foundation, Suh \textit{et al.}~\cite{suh2023sensecape} advanced knowledge comprehension by developing an interactive system that reveals multilevel abstractions of complex information spaces. Taking a different approach, Jin \textit{et al.}~\cite{jin2024teach} focused on knowledge construction by leveraging LLMs as teachable agents within a ``learning by teaching'' paradigm. 

\rr{Despite their promise, existing LLM sensemaking applications rarely consider how individual epistemological schemas shape knowledge construction.
Epistemology--the philosophical study of knowledge and belief that guides learners to evaluate the credibility and relevance of information sources--underpins how learners make sense of new information~\cite{hofer1997development, schommer2004explaining, shirzad2022epistemic}.
Pedagogical research emphasizes that the effectiveness of sensemaking depends fundamentally on learners' epistemological schemas--their structured beliefs about knowledge's \textit{nature}, \textit{source}, and \textit{justification}~\cite{hofer1997development, braaten2010personal, baker2020epistemic}. These beliefs critically shape how individuals process information during learning, from initial interpretation to final evaluation~\cite{zhang2020cognitive}. For example, Baker \textit{et al.}~\cite{baker2020epistemic} demonstrate that such personal epistemological schemas influence the entire spectrum of self-learning, affecting learners' motivation, achievement levels, and engagement with learning materials.}

\rr{Given the fundamental role of epistemology in knowledge construction, prior research has investigated applying epistemological frameworks to AI models for educational tools~\cite{woolf2013ai, zhang2023visar}.
For instance, Ros{\'e} \textit{et al.} \cite{rose2008analyzing} proposed an AI-driven framework grounded in an understanding of epistemology, which can trace learners' knowledge construction processes, thereby improving learning outcomes.
However, the rapid adoption of LLMs has shifted recent scholarly attention toward their capacity to disrupt traditional teaching practices and introduce epistemic paradoxes~\cite{sibilin2023education, chavanayarn2023navigating, cassinadri2024chatgpt}. Consequently, the potential of leveraging LLMs for accommodating learners' epistemological schemas, which is crucial for sensemaking, remains largely unexplored.}
Our research addresses this gap by integrating epistemological frameworks into LLM-based systems through prompt engineering, guiding users through the processes of problem framing, evidence evaluation, and reflection, key stages of sensemaking.
\rr{This domain-agnostic approach ensures versatility and offers a structured yet adaptable framework for knowledge exploration.}

\subsection{Toward Effective Learner-LLM Interactions}

\rr{Understanding and improving the interactivity in learning activities remains a persistent focus in pedagogy.
The \textit{Interactive}, \textit{Constructive}, \textit{Active}, and \textit{Passive} (ICAP) framework, proposed by Chi~\cite{chi2009active}, categorizes learner engagement modes based on observable behaviors and underlying cognitive processes, providing a foundation for understanding learner interactivity.
A subsequent five-year study~\cite{chi2014icap} demonstrated the relative effectiveness of these engagement modes, ranking them from most to least effective: interactive, constructive, active, and passive.
This finding has fueled further learner-centered research exploring the relationship between engagement modes and learning outcomes in specific educational scenarios~\cite{wang2015investigating, schulte2018framework}.
While some variations exist in the resulting classifications and analyses of contributing factors, the principle that ``\textit{good learning is active learning}'' remains widely accepted~\cite{brown2014make}.
However, translating the ICAP framework into practice, particularly in designing interactive and constructive activities, has proven challenging~\cite{chi2018translating}.}

\rr{Advances in technology create new opportunities to apply the ICAP framework.
Recent educational research has extended the ICAP theory to the context of TEL~\cite{sailer2021contextual, wekerle2022using, sailer2024learning}. Building on the ICAP framework, Sailer \textit{et al.}~\cite{sailer2024learning} introduced the \textit{Substitution}, \textit{Augmentation}, \textit{Modification}, and \textit{Redefinition} (SAMR) model to classify levels of technology integration in learning. The introduction of LLM-based educational agents marks a transformative step toward SAMR's ``redefinition'' level, especially in self-learning scenarios. These agents enable scalable conversational interactions that can enhance learner engagement.}

\rr{Although the conversational interface of LLM-based educational agents suggests simplicity and intuitive interaction similar to human communication \cite{smestad2019chatbot}, extant HCI literature reveals notable disparities in how learners benefit from these tools~\cite{chen2024learning}.
This variability often arises from learners' diverse motivations, levels of LLM literacy, mental models, and educational backgrounds, ultimately influencing their interaction strategies and learning outcomes~\cite{mogavi2024chatgpt, zamfirescu2023johnny}.}
For instance, research on help-seeking behaviors among self-learners highlights how self-efficacy, perceived task difficulty, and tool accessibility influence their resource utilization \cite{moores2009self, karabenick2011understanding, glassman2015overcode, urgo2022learning}. Overconfident learners, in particular, may only seek external support when facing substantial challenges or when perceiving the tool as beneficial for their understanding~\cite{moores2009self}.
\rr{This tendency can hinder their ability to accurately assess their learning progress. Furthermore, novice AI users often struggle to craft effective prompts during LLM interactions, compounding the challenges they face \cite{zamfirescu2023johnny, fiannaca2023programming}.}

\rr{To promote effective learner-LLM interaction, designing more inclusive LLM-based tools that provide effective cognitive support and accommodate the diverse needs of self-learners is essential~\cite{cohn2024towards}.
Drawing on the ICAP framework~\cite{chi2014icap} and our previously discussed epistemological perspective, our research examines how self-learners interact with LLM-based educational agents in practice. By analyzing these interaction patterns and their underlying mental models, we explore how to enhance LLM prompt engineering to better facilitate learner interactivity and knowledge co-construction.}

\section{Formative Study}


\yf{This section describes our formative study, which adopts a user-centered approach~\cite{vredenburg2002survey}. The study investigates participants’ \textbf{interaction behaviors}, \textbf{epistemological schemas}, and the \textbf{challenges} they face during self-directed learning. By synthesizing these dimensions, we derive a categorization of participants' \textit{interaction patterns} and deduce the latent \textit{mental models} that guide their engagement when learning new topics. These findings inform the design priorities of our system.}
Specifically, we outline the recruitment process and participant demographics, describe the study procedure, and explain the data analysis methods.
This study was conducted in accordance with all relevant ethical guidelines and received full approval from our university's Institutional Review Board (IRB). Prior to data collection, verbal informed consent was obtained from all participants.

\subsection{Recruitment and Participants}

\paragraph{Recruitment}
We employed a dual-pronged approach, combining convenience and snowball sampling strategies, to gather a representative sample of target users who use LLM-based chatbots for self-learning.
We disseminated recruitment messages with contact information through multiple channels, including social media, word of mouth, and campus bulletin boards at five institutions of higher education in East Asia and North America.
Our recruitment criteria consisted of the following: (1) adults aged 18 years or older, (2) individuals capable of interacting with LLM-based chatbots, (3) those with at least three months of prior experience using LLMs for self-learning, and (4) participants without prior knowledge of blockchain technology, which served as the primary reading material for our self-learning task.
We received a total of 35 applications and conducted a thorough screening of respondents based on our criteria.
We stopped the recruitment process when data analysis results reached theoretical saturation \cite{clarke2017thematic}, meaning no new insights emerged from participants' dialogue records or semi-structured interviews (SSI).
Ultimately, we recruited 26 participants with varied educational backgrounds, from various disciplines, and from seven countries and regions (see Table~\ref{tab: formative} in Appendix~\ref{apx: formative}).

\begin{table*}[tb!]
\caption{Descriptive statistics of participants in the formative study. This diverse sampling pool provides insights into design requirements for improving LLM-based chatbots for self-learning across various demographics.}
 \centering
    \includegraphics[width=\textwidth]{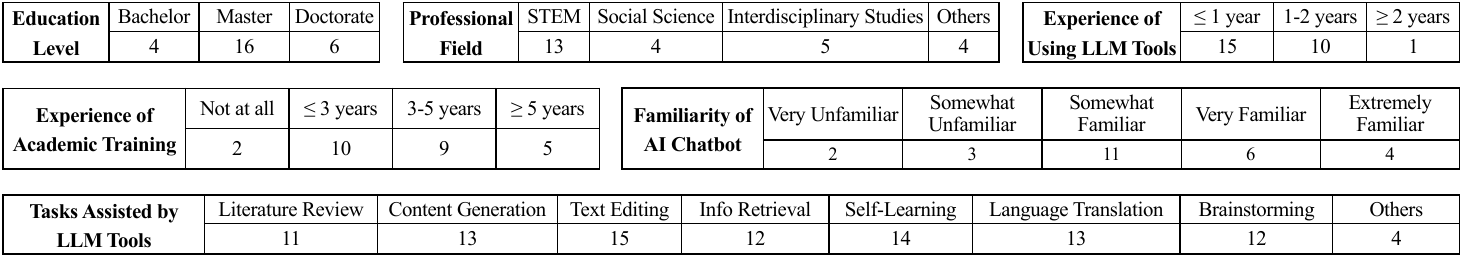}
\label{fig: form_demo}
\end{table*}

\paragraph{Participants}
The 26 participants were between 23 and 32 years of age (M = 27.04, SD = 2.28), and carefully selected to represent diverse backgrounds and experiences with LLM-based tools (see Table \ref{fig: form_demo}).
Sixteen held a master's degree, followed by four with bachelor's degrees and six with doctorates, indicating a strong representation of individuals engaged in advanced studies or research.
Our participants represented a diverse range of disciplines, with a substantial portion (13) coming from STEM fields--a group shown to have positive perceptions of LLMs as learning tools~\cite{bernabei2023students}.
The remaining participants were from social sciences (4), interdisciplinary studies (5), and other fields (4), ensuring broader perspectives.
Most participants (22) had experience in academic training, providing a foundation for evaluating LLM-based self-learning tools.
While ten participants had 1-2 years of experience using LLM tools, a significant portion (15) had been using them for less than a year, and only one for over two years, indicating a mix of early adopters and more recent users.
Finally, participants' familiarity with LLM-based chatbots varied: 21 reported being somewhat (11), very (6), or extremely (4) familiar, and five were unfamiliar.
This distribution enabled comparison between more and less adept chatbot users, a crucial aspect of our study design.

\subsection{Formative Study Design}

The formative study aimed to investigate \yf{participants' \textbf{interaction behaviors}, \textbf{epistemological schemas}, and the \textbf{challenges} they encountered} during self-learning through a task-focused probe system.
Before formally launching the formative study, a pilot study was conducted with two team members from diverse disciplinary backgrounds to validate the process and estimate the required duration.

\subsubsection{Study Setup}

The formative study allowed both in-person (n=12) and online (n=14) participation.
In-person sessions were held at the recruiting institutions. Online sessions were conducted via Zoom~\footnote{\url{https://zoom.us/}} to facilitate real-time interaction and observation of participant behavior.
To provide consistent participant experiences within a controlled learning environment, we employed Qualtrics\footnote{\url{https://www.qualtrics.com/}} for online surveys and a task-focused probe system \yf{accessed via the Web directly}.
This system included raw learning materials, a textual article, and an LLM-chatbot limited to responding only to the provided learning material (see Fig. \ref{fig: web} in Appendix \ref{apx: probweb}).
This controlled setup ensured a fair self-learning process, free from misinformation and external online sources.

Aiming for smooth operation, participants received essential documents, including the consent form and study guidance, one day prior to their scheduled session.
These documents detailed the research goals and experimental procedures and addressed any potential questions.
Notably, we did not share the probe system link in advance to avoid pre-exposure to the learning materials, which could potentially bias our assessment of user learning performance.
The formative study sessions were allocated 90 minutes based on the results of the pilot study.

\begin{figure*}[tb!]
\centering
  \includegraphics[width=\linewidth]{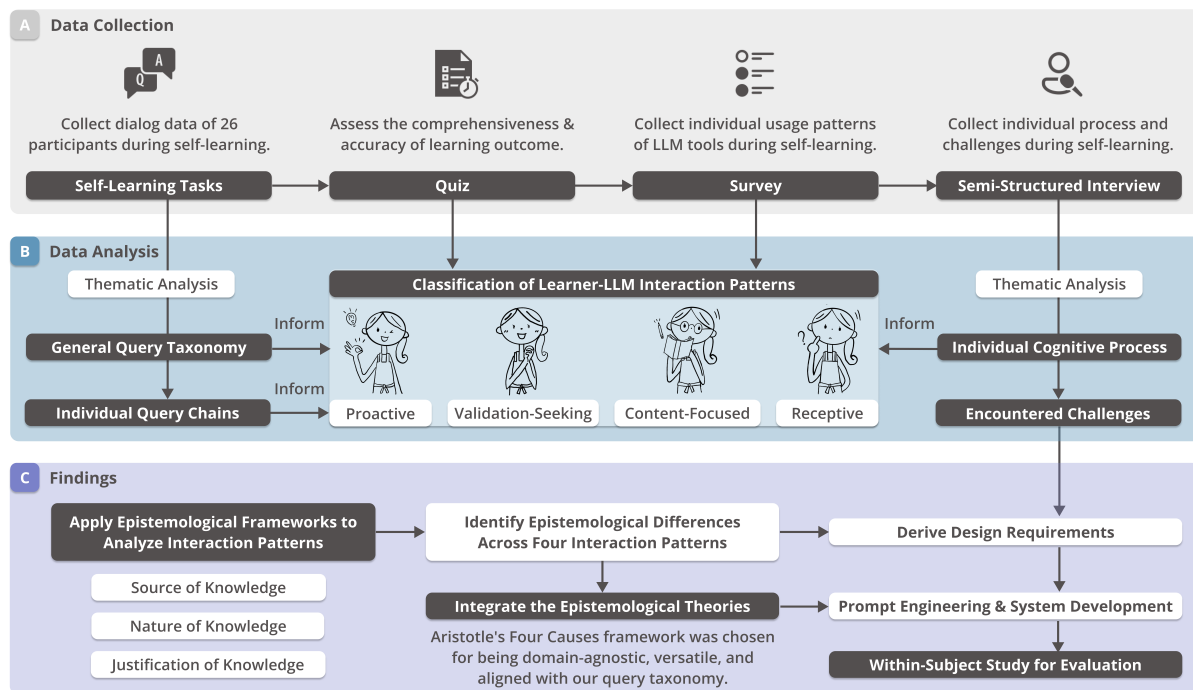}
  \caption{\rr{Formative Study Workflow (N=26): A three-phase investigation of LLM interaction patterns during self-learning. (A) \textit{Data Collection:} Gathered dialogue records, quiz scores, survey responses, and transcripts of semi-structured interviews. (B) \textit{Thematic Analysis:} Classified four distinct LLM interaction patterns (proactive, validation-seeking, content-focused, receptive). (C) \textit{Epistemological Analysis and Design Implications:} Compared and integrated epistemological frameworks across the identified interaction patterns, revealing insights into participant challenges and informing the design requirements for an epistemologically-grounded LLM-assisted learning system.}} 
  \label{fig: workflow}
\end{figure*}

\subsubsection{Study Procedure}

Upon receiving informed consent, the study proceeded with the following steps (Fig.~\ref{fig: workflow}A):

\paragraph{Self-Learning Task}
Participants initially completed a sensemaking task using an informational article on non-fungible tokens (NFTs) \cite{NFTex}.
Using a think-aloud protocol, participants provided real-time feedback on the probe system's design and usability during their interactions.
Participants self-regulated their learning pace, ceasing interaction once they felt they had adequately understood the article.
They then downloaded a record of their dialogue with the LLM for future reference.
Task completion time ranged from 15 to 35 minutes (M = 20.21, SD = 5.12).

\paragraph{Quiz}
Participants then completed a 15-minute, ten-point quiz assessing the comprehensiveness and accuracy of their acquired knowledge. The quiz, consisting of four multiple-choice and two short-answer questions, was designed by two NFT domain experts who also independently graded the responses.
The final scores were the average of the two experts' grades.

\paragraph{Survey and Semi-Structured Interview}
To further understand individual learning experiences, a seven-point Likert scale survey was administered, gathering data on LLM usage patterns (e.g., familiarity, modalities) and perceived challenges.
Semi-structured interviews then explored individual learning processes in depth, with participants sharing the rationale behind their query chains, challenges, and suggestions for improving LLM interactions.~\footnote{\rr{For illustrative examples of conversation records corresponding to each interaction pattern, please refer to the supplementary materials.}}
Each interview lasted between thirty minutes and one hour, with audio recordings and verbatim transcriptions conducted.

\begin{figure}[tb!]
\centering
  \includegraphics[width=\linewidth]{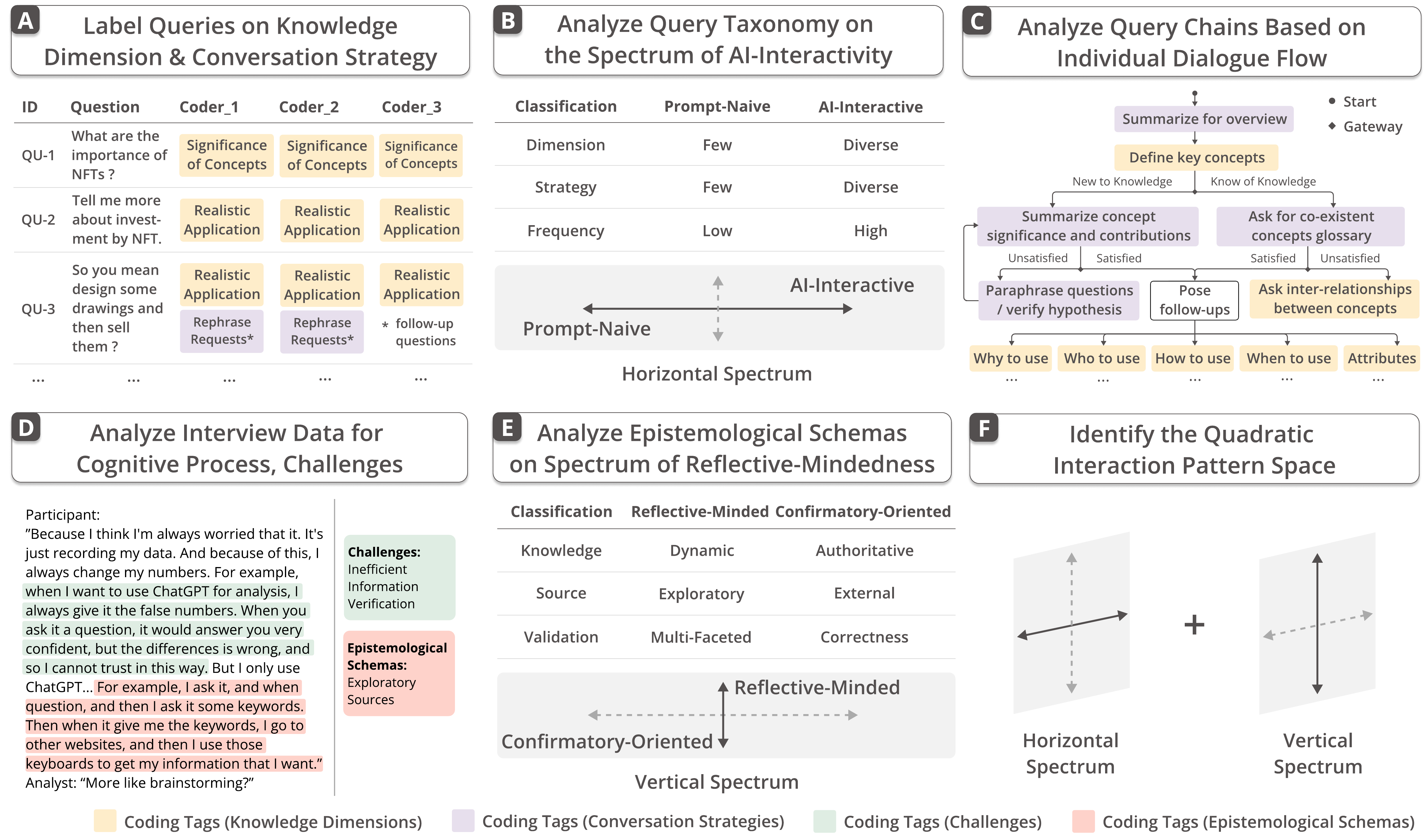}
  \caption{\rr{This figure presents our six-step process for analyzing participants' interaction behaviors and epistemological frameworks based on data collected from our formative study.}} 
  \label{fig: l_pattern}
\end{figure}

\subsection{Data Analysis}

\rr{We employed reflexive thematic analysis~\cite{braun2019reflecting, braun2021one} to explore participants' sensemaking processes and categorize their interaction patterns.
Building on existing literature~\cite{aliannejadi2021analysing, schneider2023investigating}, we defined interaction patterns based on two key aspects. The first focuses on participants' \textbf{interaction behaviors}, as reflected in the queries they generated during the self-learning process. 
\yf{The second dimension pertains to their \textbf{epistemological schemas}, derived from their chain of queries and self-reported knowledge construction processes detailed in the interview transcripts.}
We also analyzed the \textbf{challenges} participants encountered while interacting with the LLM through a detailed examination of the interview transcripts, which was further complemented by their survey responses. The complete process and use of collected data sources are illustrated in Fig. \ref{fig: workflow}B.

Given the exploratory nature of our study within the emerging field of LLM-mediated self-learning, we adopted a predominantly \textit{inductive} approach.
This approach enabled us to develop a query taxonomy of learner-LLM interactions while drawing on an epistemological framework~\cite{aksan2009descriptive, mokhtari2014quantitative, huang2023students} to analyze participants' knowledge construction processes.
This flexible approach enabled us to explore the multifaceted nature of user experiences in depth~\cite{guest2011applied}, while grounding our analysis in relevant theoretical constructs.}
Three analysts participated: two authors (who observed all self-learning tasks and conducted all interviews) and one NFT domain expert.

\subsubsection{Interaction Behaviors Analysis}
\yf{Initially, the three analysts independently examined participants' interaction behaviors within the self-learning task (see Fig. \ref{fig: l_pattern}A).}
The goal was to systematically categorize 179 participant-generated queries based on the \textit{knowledge dimensions} explored and the \textit{conversation strategies} employed. The analysts met regularly to share coding results, discuss discrepancies, and refine the codebook until reaching a consensus for the query taxonomy.
The resulting query taxonomy not only highlights the types of queries generated by participants but also captures the frequency of their interactions and the diversity of conversation strategies they employed.
\yf{Based on this analysis, participants’ interaction behaviors were mapped along a
spectrum ranging} from \textbf{prompt-naive} (low interaction frequency; few exploration dimensions and conversation strategies) to \textbf{AI-interactive} (high interaction frequency; diverse exploration dimensions and conversation strategies) (see Fig. \ref{fig: l_pattern}B).
To further contextualize these patterns, the analysts conducted a detailed examination of each participant's chain of queries (see Fig. \ref{fig: l_pattern}C). This examination informed the subsequent semi-structured interviews designed to elicit a nuanced understanding of participants' epistemological schemas for self-learning.

\subsubsection{Epistemological Schemas Analysis}
\label{sec: epiframe}
\rr{To investigate participants' epistemological schemas, we integrated the previously established query taxonomy, the analysis of query chains, and the data gathered from the semi-structured interviews. During these interviews, we asked participants about their information needs and how they constructed meaning, referencing their specific query flows from the dialogue records.
We analyzed the interview data (22.5 hours recorded, transcribed into 26 transcripts) alongside the corresponding dialogue records.
This analysis employed a three-dimensional framework focusing on the nature, source, and justification of knowledge to understand participants' personal epistemological schemas~\cite{aksan2009descriptive, mokhtari2014quantitative, huang2023students}.
Guided by this framework, we systematically coded the interview data within \textit{Atlas.ti}~\footnote{https://atlasti.com/} software (see Fig. \ref{fig: l_pattern}D).
We met weekly to review findings, resolve disagreements, and refine the codes until a consensus was reached.
As a result, we placed participants' epistemological schemas along a spectrum from \textbf{reflective-minded} to \textbf{confirmatory-oriented} (see Fig. \ref{fig: l_pattern}E). Reflective-minded patterns are characterized by a dynamic view of knowledge, knowledge construction through exploration, and justifications through multi-faceted validation. In contrast, confirmatory-oriented patterns demonstrate an authoritative view of knowledge, reliance on external expertise, and justification by seeking standard answers.}

\subsubsection{Interaction Patterns Classification}
\rr{This mixed-methods approach revealed two key spectra for categorizing participants based on a two-dimensional analytical framework: \textbf{interaction behavior}, ranging from prompt-naive to AI-interactive, and \textbf{epistemological schema}, spanning from reflective-minded to confirmatory-oriented. The integration of these dimensions provided a robust and holistic analysis, generating a quadratic classification of interaction patterns (similar to~\cite{he2023enthusiastic}): \textit{proactive}, \textit{validation-seeking}, \textit{content-focused}, and \textit{receptive} (see Fig. \ref{fig: l_pattern}F).}


\begin{figure}[tb!]
\centering
  \includegraphics[width=0.65\linewidth]{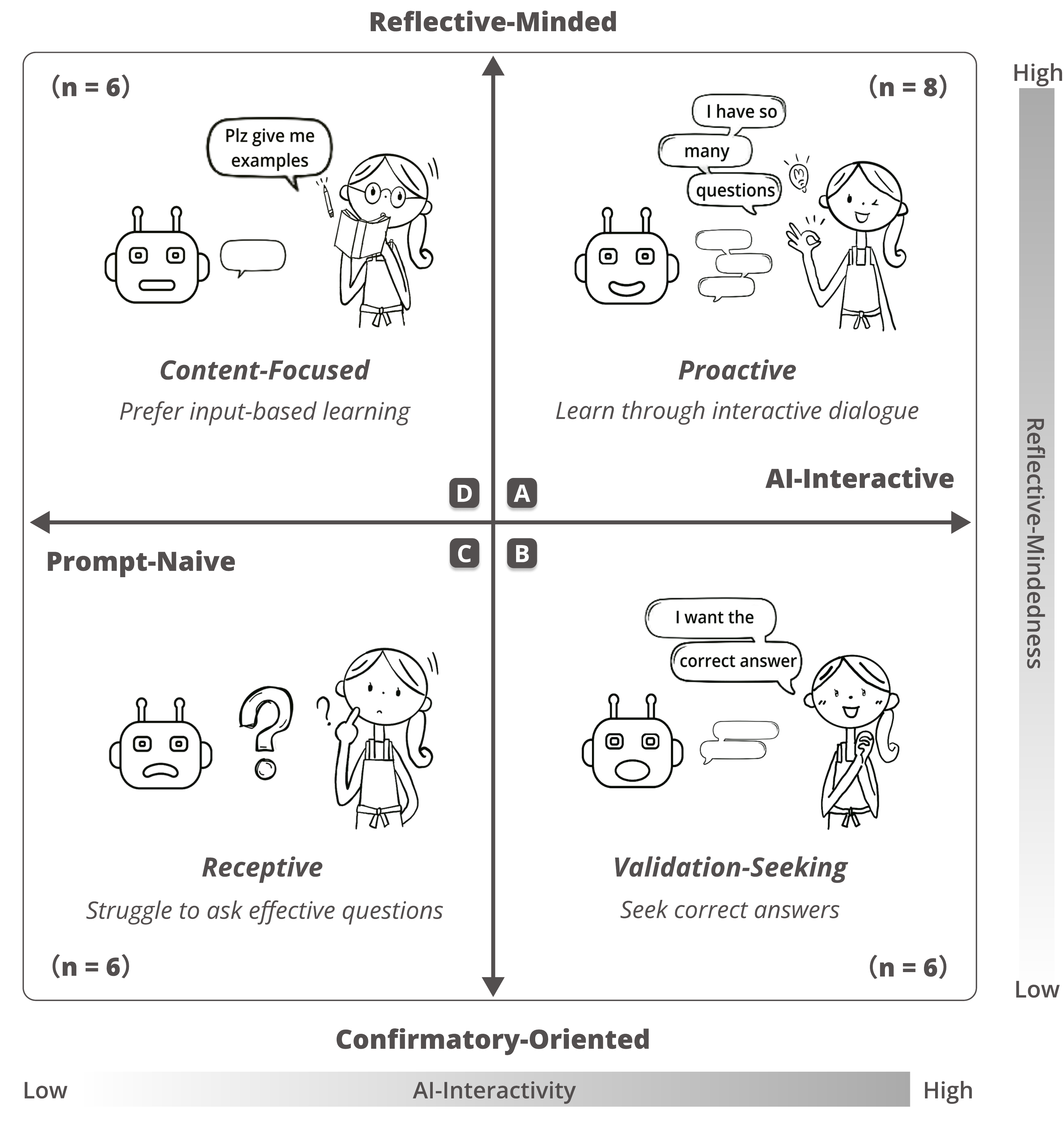}
  \caption{\rr{This 2x2 matrix presents four distinct interaction patterns, categorized by AI interactivity and reflective-mindedness: Prompt-Naive, Confirmatory-Oriented, Reflective-Minded, and AI-Interactive. This classification reflects the mental models associated with different learning strategies observed during self-learning.}} 
  \label{fig: qua_class}
\end{figure}

\section{Findings}

This section \yf{unpacks four interaction patterns and their epistemological differences.}
These analyses, combined with survey responses and interview data, revealed key challenges and informed the design requirements for our LLM-supported self-learning prototype (see Fig. \ref{fig: workflow}C).

\subsection{Learner-LLM Interaction Patterns}

This section illustrates four learner-LLM interaction patterns.
\rr{These patterns are characterized by their query types, conversational strategies, self-reported learning workflows (gathered through semi-structured interviews), and quiz performance.}

\subsubsection{Proactive (n = 8)}

\rr{Eight participants exhibited the proactive interaction pattern marked by high interactivity and reflectiveness and actively pursued knowledge} (see Fig. \ref{fig: qua_class}A).
They viewed the chatbot as a collaborative partner, engaged in extended, multi-round dialogues, and critically examined responses.

\yf{The proactive interaction pattern is demonstrated through the participants’ elevated query volume and diverse exploratory query types (see Table \ref{table: query_type}).}
They delved deeper into topics by asking about ``attributes of concepts,'' ``co-existent concepts,'' ``realistic application,'' and ``real-world consequences''--all highlighting a desire to understand a topic from multiple angles.
For example, P9 explained, ``\textit{I don't just want the `what,' I want the `why' and the `how does this connect to other things?'}''

In addition, these participants strategically employed queries such as ``change perspectives'' and ``rephrase requests,'' actively guiding the conversation toward providing a more holistic understanding.
For instance, P12 revealed, ``\textit{I rephrase and ask about the same concept multiple times in different ways to get a more comprehensive and specific response, instead of an overly general one.}''
This demonstrates an active role in shaping the learning experience.
This interaction pattern was often reflected in their high quiz scores, ranging from 7 to 9 out of 10 (M = 8.28, SD = 0.62), and suggested active learning strategies driven by curiosity and a desire for deeper comprehension.

\begin{table*}[tb!]
\caption{\rr{This table presents the query taxonomy developed during the formative study, structured around three key aspects: \textit{Knowledge Dimension}, \textit{Conversation Strategy}, and \textit{Interaction Frequency} (visualized numerically via a heat map).} Questions marked with an asterisk (*) represent follow-up queries designed to provide deeper insights. While the heat map directly quantifies the AI-interactivity dimension, reflective-mindedness—captured through triangulated interview analysis—is only partially represented here.}
 \centering
    \includegraphics[width=\textwidth]{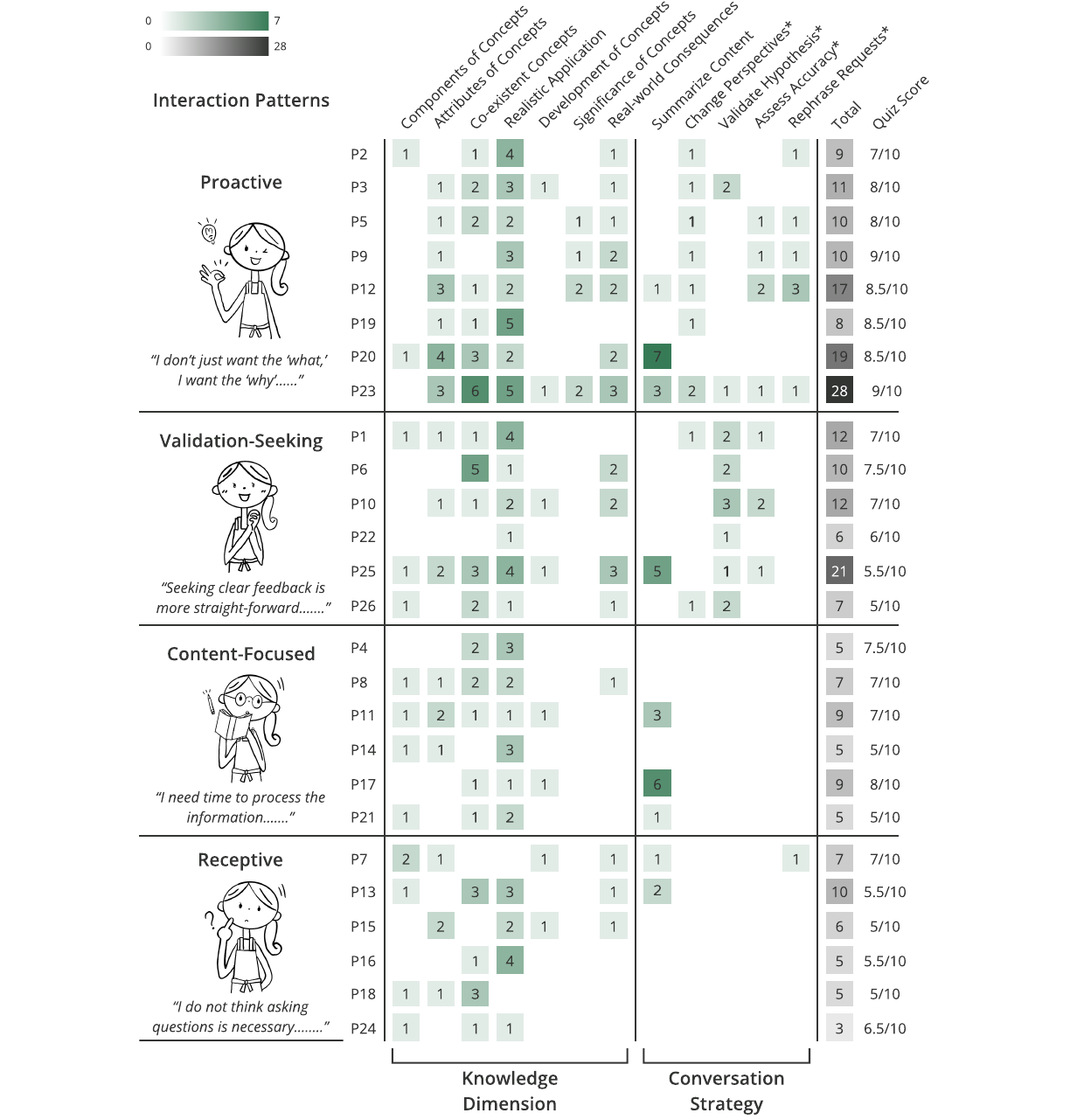}
\label{table: query_type}
\end{table*}

\subsubsection{Validation-Seeking (n = 6)}

\rr{Six participants displayed the validation-seeking interaction pattern, characterized by high interactivity but low reflectiveness, primarily engaging with chatbots to confirm existing knowledge} (see Fig. \ref{fig: qua_class}B).
\yf{Their high-frequency interaction behaviors primarily involve asking specific questions with predetermined expectations to validate assumptions or test hypotheses (see ``validate hypothesis'' in Table \ref{table: query_type})}
Their repeated queries about ``co-existent concepts'' and ``realistic applications'' reflected this desire to align new information with established understanding and real-world scenarios (see Table \ref{table: query_type}). 
As P1 noted, ``\textit{Relating new concepts to hypotheses formed from my existing knowledge boosts my confidence in self-learning.}''

Furthermore, they often apply conversational strategies such as ``validate hypothesis'' and ``assess accuracy,'' emphasizing their preference for reinforcing existing knowledge rather than exploring new perspectives.
As P22 shared, ``\textit{Instead of constantly trying to synthesize vast new concepts, presenting a hypothesis and seeking clear feedback is more straightforward.}''
Their quiz scores, ranging from 5 to 7.5 out of 10 (M = 6.33, SD = 0.98), suggest that while these participants prioritize accuracy, relying on their existing understanding may not always facilitate the necessary expansion of their knowledge base for a thorough grasp of the subject.

\subsubsection{Content-Focused (n = 6)}

\yf{Six participants demonstrated the content-focused interaction pattern, characterized by low interactivity but high reflectiveness in their behaviors, prioritizing thorough understanding over fragmented knowledge}
(see Fig. \ref{fig: qua_class}D).
These participants often engaged in extensive exploration of learning materials before directly engaging with the LLM chatbot. As P14 articulated, ``\textit{I do not want to ask superficial questions.
I need time to process the information and figure out valuable questions}.''
With this pattern, participants tended to use the LLM less often, primarily for retrieving information directly related to specific aspects of the learning materials, such as ``components of concepts,'' ``co-existent concepts,'' and ``realistic applications.''

This pattern is further evidenced by their sole use of the ``summarize content'' strategy (see Table \ref{table: query_type}).
P17 aptly captured this tension, noting, ``\textit{Simultaneously grasping new concepts and forming insightful questions is challenging, so I mainly use the LLM to facilitate understanding}.''
However, this approach may not always translate to high performance within a time-limited assessment context, as suggested by the significant individual variability in quiz scores, ranging from 5 to 8 out of 10 (M = 6.58, SD = 1.64).

\subsubsection{Receptive (n = 6)}

\yf{The receptive interaction pattern was evident in six participants, whose behaviors were marked by minimal engagement, limited reflectiveness, and passive interaction with the LLM chatbot}
(see Fig. \ref{fig: qua_class}C). 
Accustomed to traditional learning methods like reading, online courses, and structured video lectures--which typically involve limited interaction--these participants often struggled to formulate precise questions or to fully leverage the chatbot's capabilities.
As P13 admitted, ``\textit{I am not used to learning with an LLM. I do not think asking questions is necessary.}'' This sentiment is further evidenced by a lack of discernible patterns in their query topic and strategy preferences, indicating a less systematic LLM-interaction approach (see Table \ref{table: query_type}).
Furthermore, their quiz scores, ranging from 5 to 7 out of 10 (M = 5.75, SD = 0.82), generally suggest that those who struggle to adapt to this interactive learning modality may experience less favorable learning outcomes.

\subsection{User-Centered Design}
\label{sec: dr}

\rr{This section illuminates the epistemological differences across the four learner-LLM interaction patterns. Integrating insights from the surveys and interviews,} this analysis reveals the primary challenges (\textbf{C1-3}) participants face when using LLMs and informs design requirements (\textbf{DR1-5}) for enhancing LLM-supported self-learning tools.

\begin{table*}[htb!]
\caption{\rr{This table presents the matrix analysis of epistemological differences across four interaction patterns based on the three-dimensional epistemological framework.}}
 \centering
    \includegraphics[width=\textwidth]{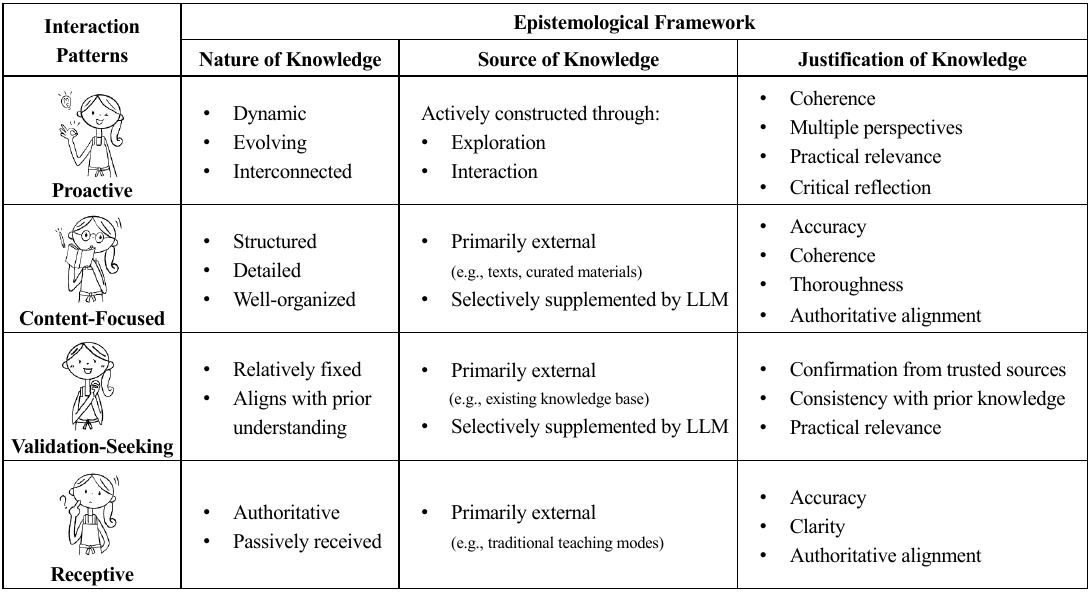}
\label{table: epi_gap}
\end{table*}

\subsubsection{Epistemological Differences in LLM Interactions}

\rr{A comparative analysis of the four interaction patterns, presented in Table~\ref{table: epi_gap} using a matrix method~\cite{groenland2018employing}, reveals notable epistemological differences.
This analysis, informed by the previously established three-dimensional epistemological framework~\cite{hofer1997development, aksan2009descriptive, mokhtari2014quantitative, huang2023students} described in \S\ref{sec: epiframe}, considers the perceived \textit{nature} of knowledge (absolute vs. evolving), the primary \textit{source} of knowledge (external authorities vs. active construction), and the criteria for \textit{justifying} knowledge.}

\rr{The contrast between proactive and receptive interaction patterns revealed fundamentally different mental models of sensemaking.
Proactive patterns, marked by actively seeking diverse perspectives and critical reflection, facilitate dynamic knowledge construction and validation through iterative interaction and exploration.
While this iterative exploration promotes deep understanding, it creates a practical challenge of efficiently verifying the extensive information generated through LLM interactions (see \textbf{C1} in \S\ref{sec:challenge-in-llm}).
In contrast, receptive patterns rely heavily on external authority and a more passive way of engaging, which limits effective knowledge construction with LLMs. As a result, fostering meaningful interactivity and learner agency (see \textbf{C2} in \S\ref{sec:challenge-in-llm}) becomes particularly challenging when users treat LLMs primarily as information sources rather than interactive learning partners.}

\rr{Participants exhibiting validation-seeking and content-focused interaction patterns occupy an interesting middle ground.}
While they value accuracy and structure, their strategic use of LLMs for targeted information suggests an openness to expanding their understanding beyond curated materials.
\rr{However, these patterns tend to emphasize alignment with authoritative sources and validation of prior knowledge. They may inadvertently limit learners' exploration of diverse perspectives and create a gap between their desire for comprehension and reliance on established sources (see \textbf{C3} in \S\ref{sec:challenge-in-llm}).}

\subsubsection{Challenges in LLM Interactions}
\label{sec:challenge-in-llm}

\begin{figure}[tb!]
\centering
  \includegraphics[width=0.99\linewidth]{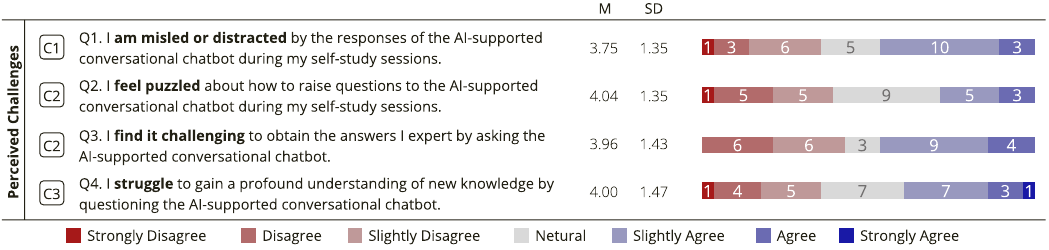}
  \caption{\rr{A substantial portion of participants reported challenges using the LLM-supported chatbot for self-study, particularly with question formulation, response clarity, and the achievement of deep understanding.}} 
  \label{fig: challenge}
\end{figure}

Our analysis of epistemological differences across different LLM interaction patterns, integrating survey (see Fig. \ref{fig: challenge}) and semi-structured interview results, reveals unique challenges participants encountered during LLM-mediated self-learning.

\begin{enumerate}[leftmargin=*, label=\textbf{C\arabic*}.]

\item 
\textbf{Inefficient Information Verification.}
Across various \rr{interaction patterns}, effectively and efficiently verifying the accuracy of LLM-generated information poses a significant challenge~\cite{cucuiat2024feedback}.
\rr{Participants with proactive and validation-seeking interaction patterns find the validation process time-consuming and cumbersome, hindering their ability to assess the reliability of AI's responses.
For instance, P9 commented, ``\textit{ChatGPT sometimes fabricates information, so I have to verify its explanations against multiple sources, which can takes longer than finding the answer myself.}''}

\item
\textbf{Limited Interactivity and Agency.}
The interactive nature of LLMs presents challenges to self-learners accustomed to more structured or passive approaches~\cite{zamfirescu2023johnny}.
\rr{Participants showing content-focused interaction patterns,} find that the lack of seamless integration between LLM responses and their preferred learning materials diminishes their sense of control over the learning process.
\rr{As P11 noted, ``\textit{Switching between the chat and readings is frustrating. Especially when the AI's responses do not align with my learning goals--it just throws me off track.}''}
Similarly, \rr{participants with receptive tendencies} struggle to adapt to the open-ended nature of interacting with LLMs, finding it difficult to formulate effective queries and manage the volume of information received.
\rr{As P24 explained, ``\textit{Asking good questions is challenging in itself--you need to know the topic already. Helpful prompts from the AI would be great; otherwise, I am lost.}''}

\item
\textbf{Confirmation Bias.}
\rr{Participants showing validation-seeking preferences are particularly vulnerable to confirmation bias.}
They use LLMs to reinforce pre-existing beliefs rather than critically examining alternative perspectives or challenging their own assumptions.
\rr{For example, P22 admitted, ``\textit{I sometimes keep tweaking my prompts until ChatGPT gives me what I expected, even if that means ignoring other perspectives.}''
This behavior underscores how unstructured LLM interactions can narrow learning pathways. It highlights the need for systematic query approaches and metacognitive support to help learners examine their thinking processes and develop more comprehensive exploration strategies.}

\end{enumerate}

\subsubsection{Design Requirements}

The challenges uncovered in our formative study directly informed the five key design requirements for our LLM-supported self-learning system:

\begin{enumerate}[leftmargin=*, label=\textbf{DR\arabic*}.]

\item
\textbf{Validated Information and Validation Support (\rr{C1}).}
Systems must provide readily validated information or offer user-friendly methods for verifying the accuracy of LLM outputs.
This includes facilitating precise cross-referencing between LLM responses and the original source material.

\item
\textbf{Smooth Onboarding and Reduced Cognitive Load (\rr{C2}).}
Learners expressed a preference for LLMs to autonomously generate systematic follow-up questions, minimizing the initial interaction effort. This approach reduces the barrier to entry and enhances self-study efficiency by reducing the time spent crafting prompts.

\item
\textbf{Sustained Engagement and Deepened Exploration (\rr{C2}).}
Engaging and fluid dialogue is crucial to maintaining users' engagement while guiding them toward a deeper understanding of new concepts. Supporting flexible, multi-round dialogues enables learners to progressively deepen their understanding of concepts through both horizontal and vertical exploration.

\item
\textbf{Fostering Openness and Multidimensional Thinking (\rr{C3}).}
Comprehensive knowledge acquisition requires learners to consider multiple perspectives and generate diverse hypotheses.
A well-designed taxonomy of queries, informed by a robust epistemological framework, can effectively support this type of multifaceted thinking.

\item
\textbf{Facilitating Metacognition and Reflection (\rr{C3}).}
Systems should support metacognition by structuring the interaction process and encouraging reflection on learning progress and comprehension levels.
Recording and visualizing the user's chain-of-query can enhance knowledge retrieval efficiency and help identify potential gaps in understanding.

\end{enumerate}

\section{CausaDisco}

We present \textit{CausaDisco}, an epistemologically-informed system designed to enhance sensemaking during self-learning.
Carefully designed features enable \textit{CausaDisco} to encourage user interaction with an LLM chatbot to enhance sensemaking, fulfilling the design requirements outlined in \S\ref{sec: dr}.
In the following sections, we illustrate \textit{CausaDisco}'s interface design, highlighting its integration of the epistemological framework and prompt engineering techniques.

\subsection{Interface Design}

\begin{figure*}[tb!]
\centering
  \includegraphics[width=\linewidth]{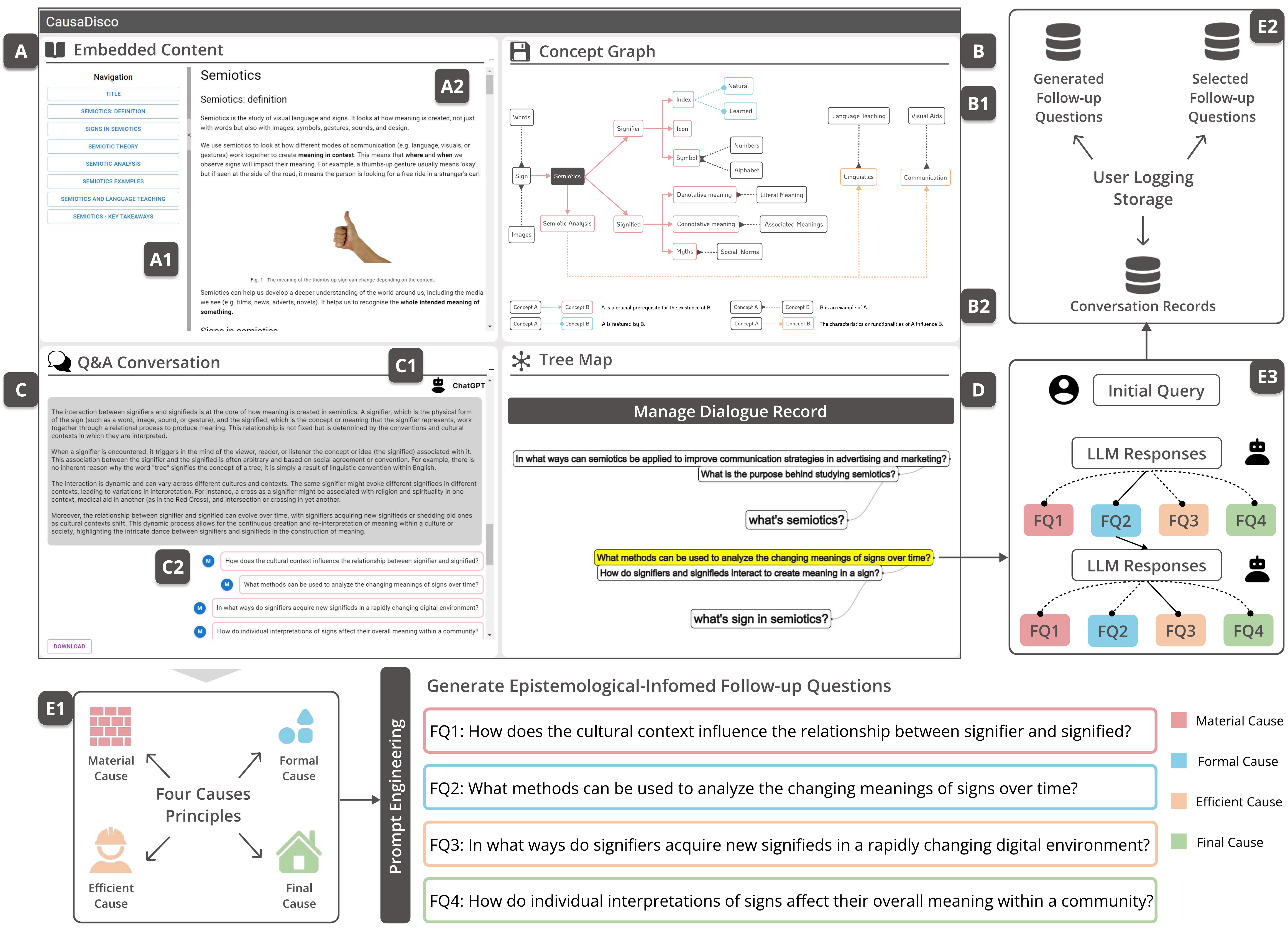}
  \caption{System Overview: The interface features four main views: A) \textit{Embedded Content View}; B) \textit{Concept Graph View}; C) \textit{Q\&A Conversation View}; D) \textit{Tree Map View}. 
 The backend comprises two main components: (E1) The Four Causes principles, which inform the prompt engineering for generating follow-up questions (FQ); and (E2) A logging function that records the user's chain of queries (E3) during interactions with the LLM chatbot.} 
  \label{fig: interface}
\end{figure*}

\textit{CausaDisco} integrates four coordinated views to streamline the self-learning process within complex educational content. 
The \textit{Embedded Content View} (Fig. \ref{fig: interface}A) provides users with access to the original learning materials, allowing immediate verification and reference (\textbf{DR1}).
Complementing this, the \textit{Concept Graph View} (Fig. \ref{fig: interface}B) visually maps out the core concepts and their interrelationships, clearly and concisely representing the structure of the material (\textbf{DR2}). 
The interactive \textit{Q\&A Conversation View} (Fig. \ref{fig: interface}C) enables users to interact with the chatbot and to choose from a range of suggested follow-up questions, fostering a deeper exploration of specific topics (\textbf{DR3, DR4}). 
Lastly, the \textit{Tree Map View} (Fig. \ref{fig: interface}D) organizes the user's inquiry process into a logical framework, facilitating easy retrospection and effectively tracking their learning progression (\textbf{DR5}).

\subsubsection{Embedded Content View}

The \textit{Embedded Content View} (Fig. \ref{fig: interface}A) has two sections. 
The right-hand side displays the original learning materials, including hierarchical headings, author information, publication date, main text, and illustrations (when available) (Fig. \ref{fig: interface}A2).
The left-hand side provides an interactive navigation bar for users to locate sections of interest swiftly and can be hidden to maximize screen space (Fig. \ref{fig: interface}A1).
This design facilitates users in verifying the authenticity or accuracy of LLM responses (\textbf{DR1}). 
They can effortlessly use the navigation bar to pinpoint the pertinent section in the original text to cross-reference information and delve into more details.

\paragraph{Justifications:}
The design of the \textit{Embedded Content View} prioritizes text-based learning resources, as informed by our formative study findings.
This focus on text enables seamless AI integration without disrupting learning flow while allowing efficient content parsing for accurate, context-aware AI responses. Text-based content also offers flexible presentation and interaction, accommodating diverse \rr{interaction patterns}.

\subsubsection{Concept Graph View}
\label{subsec: conceptV}

The \textit{Concept Graph View} (Fig. \ref{fig: interface}B) comprehensively summarizes the core concepts and their interrelationships within the learning materials.
We employ a semi-automated approach to enhance the accuracy and granularity of the concept graph.
By first identifying high-frequency keywords in the materials, we established the scope of the concept graph by incorporating a glossary of concepts provided by ChatGPT-4.
After analyzing the original texts with assistance from two domain experts in the respective fields (in our case, NFTs and semiotics), we identified four types of relationships between the key concepts: foundational prerequisites, defining traits, illustrative examples, and the influence of one concept's functions or characteristics on another (Fig. \ref{fig: interface}B2).
Accordingly, we created the corresponding concept graphs for different learning materials (Fig. \ref{fig: interface}B1).
This view offers users a concise overview of the knowledge landscape, using keywords as search indices to create a straightforward starting point for interacting with the LLM (\textbf{DR2}).


\subsubsection{Q\&A Conversation View}
\label{subsec: conversationV}

The \textit{Q\&A Conversation View} (Fig. \ref{fig: interface}C) equips users with epistemologically-informed follow-up questions during their interactions with the LLM chatbot. 
This view autonomously generates four alternative follow-up questions immediately after the chatbot responds to the previous query (Fig. \ref{fig: interface}C2).
These follow-up questions support the deep exploration and comprehension of the concepts discussed, enriching the experience beyond the typical single-turn conversation (\textbf{DR3}).  
Users can interact with any of the automatically generated follow-up questions by clicking on them to obtain a response from the LLM chatbot. 
They can also customize a question by clicking the ``modify'' button next to it. 
This button allows them to edit the question in the input box before submitting it.

The follow-up questions are crafted to guide users in exploring the topic from multiple perspectives (\textbf{DR4}), fostering a more holistic understanding. The underlying mechanism for generating these questions is discussed in detail in \S\ref{sec: prompt}. 
This view also logs all dialogue interactions between users and the LLM chatbot (Fig. \ref{fig: interface}E2).

\subsubsection{Tree Map View}

The \textit{Tree Map View} (Fig. \ref{fig: interface}D) dynamically visualizes users' query histories, allowing for a quick review of their learning progress and facilitating timely adjustments to their learning strategies (\textbf{DR5}).
This view is characterized by its cross-view interactions. When users interact in the \textit{Q\&A Conversational View}, the \textit{Tree Map View} simultaneously and automatically records the ``parent'' questions they ask. 
In a hierarchical node-link format, it presents the ``child'' follow-up questions selected by users (Fig. \ref{fig: interface}E3). 
The query tree guides users to progressively explore the topic in depth until a new question is posed, indicating the end of the learning session for the previous parent topic.
At this point, the view automatically creates a new tree map for the new parent topic. 

Users can freely expand or collapse branches associated with the same parent topic to adjust the level of detail shown in the tree map. 
They can also adjust the tree map's size and position through scrolling and panning. 
This view preserves all generated tree maps, assisting users in managing their interaction records and reviewing their logical structures, supporting reflection on the learning process, and adjusting their question strategies.

\paragraph{Justifications:} 

We also considered using a table format to display users' query histories. 
However, feedback from participants during our formative study suggested that graphs are more intuitive than tables, as they better illustrate the hierarchical relationships between questions.


\subsection{Prompt Engineering}
\label{sec: prompt}

This section details the process of generating the follow-up questions outlined in \S\ref{subsec: conversationV}. We discuss the epistemological framework we selected, the rationale behind its choice, and how it is incorporated into our prototype.

\subsubsection{Leveraging the Epistemological Framework}

As discussed in \S\ref{sec: dr}, we aim to design follow-up questions that facilitate both comprehensive exploration and multidimensional thinking. Assigning the LLM, ChatGPT in our case, the role of a tutor seems appropriate; however, this poses the challenge that the generated questions might merely be relevant without addressing the user's understanding gaps. Furthermore, ChatGPT's tendency to mimic expert vocabulary~\cite{zollman2023analyzing} may be unsuitable for our application scenario. Therefore, we equip ChatGPT with a \rr{structured framework} to guide its output, ensuring the follow-up questions provide effective cognitive support for self-learners.

Given that the follow-up questions are designed to facilitate a comprehensive exploration of the topic, incorporating this \rr{framework} into the prompt should yield questions that span all categories we summarize in Table~\ref{table: query_type}.
Furthermore, our approach needs to be domain-agnostic and easily integrated into prompts.
Consequently, we opted for an established theory that ChatGPT understands and incorporated it as a guiding term in our prompts.

After exploring various options, we found that popular learning theories from behaviorism or cognitivism were inadequate for our needs.
Additionally, studies on student mental models often focus on specific domains \cite{fratiwi2020developing,taylor2003promoting}. 
To address this, we turned to epistemology, which centers on acquiring knowledge and understanding reality. Epistemological theories have long been integrated into pedagogical practices to foster critical thinking and active engagement in exploring the unknown \cite{duschl1990psychology, macallister2012virtue, kotzee2018applied}.
Among various epistemological theories, Aristotle's \textit{Four Causes} framework (\textit{Material}, \textit{Formal}, \textit{Efficient}, and \textit{Final} Causes)~\cite{hocutt1974aristotle} \rr{stands out for its ability to refine cognitive processes~\cite{perez2007aristotle} and improve educational outcomes~\cite{taylor2023using}.}
Specifically, this framework offers a comprehensive analytical lens for understanding the essence of any entity, covering all question types summarized in Table~\ref{table: query_type}.

We used a deductive method to map the connections between our query taxonomy and the \textit{Four Causes} framework (Table~\ref{tab: codebook}). The material cause naturally corresponds with the ``components of concepts'' category, which refers to the fundamental elements constituting a concept. The formal cause, which describes the distinct form or arrangement defining an object, aligns with the ``attributes of concepts'' category. Additionally, the formal cause includes ``co-existent concepts,'' helping clarify the characteristics that distinguish a concept from others. The efficient cause, representing the driving force behind an entity's existence or change, aligns with both ``realistic application'' and ``development of concepts,'' as these categories explore the motivations and mechanisms of change. Lastly, the final cause, which denotes the purpose or intended outcome of an entity, maps to the ``significance of concepts'' and ``real-world consequences'' categories \cite{falcon2006aristotle,cohen2000aristotle,charlton1983aristotle}.

\begin{table*}[tb!]
\caption{The mapping between the epistemological framework, i.e., \textit{Four Causes}, and the query taxonomy identified from the dialogue data generated by participants during the self-learning task.}
\label{tab: codebook}
\resizebox{\textwidth}{!}{%
\begin{tabular}{|l|l|l|l|}
\hline
\textbf{Four Causes} & \textbf{Original Definition} & \textbf{Query Types} & \textbf{Examples} \\ \hline
\textbf{Material Cause} & \begin{tabular}[c]{@{}l@{}}``\textit{That out of which a thing comes} \\ \textit{to be and which persists.}''\end{tabular} & \begin{tabular}[c]{@{}l@{}}Components \\ of Concepts\end{tabular} & \textit{\begin{tabular}[c]{@{}l@{}}``\textit{Why must NFTs be involved with blockchain?}''\\ ``\textit{Can users mint NFTs without any assets?}''\end{tabular}} \\ \hline
\multirow{2}{*}{\textbf{Formal Cause}} & \multirow{2}{*}{\begin{tabular}[c]{@{}l@{}}``\textit{The form or the archetype,} \\ \textit{i.e. the statement of the essence.}''\end{tabular}} & \begin{tabular}[c]{@{}l@{}}Attributes \\ of Concepts\end{tabular} & \textit{\begin{tabular}[c]{@{}l@{}}``\textit{What is the meaning of non-fungible?}''\\ ``\textit{Do NFTs protect copyrights?}''\end{tabular}} \\ \cline{3-4} 
 &  & \begin{tabular}[c]{@{}l@{}}Co-existent\\ Concepts\end{tabular} & \textit{\begin{tabular}[c]{@{}l@{}}``\textit{What are the benefits of NFTs compared to cryptocurrencies?}''\\ ``\textit{What are the relationships between NFTs, ETH, and blockchains?}''\end{tabular}} \\ \hline
\multirow{2}{*}{\textbf{Efficient Cause}} & \multirow{2}{*}{\begin{tabular}[c]{@{}l@{}}``\textit{The primary source of change and} \\ \textit{the various factors that contribute} \\ \textit{to that change.}''\end{tabular}} & \begin{tabular}[c]{@{}l@{}}Realistic\\ Application\end{tabular} & \textit{\begin{tabular}[c]{@{}l@{}}``\textit{How are Non-fungible tokens (NFTs) regulated?}''\\ ``\textit{What are the uses of NFTs beyond investigation and collection?}''\end{tabular}} \\ \cline{3-4} 
 &  & \begin{tabular}[c]{@{}l@{}}Development \\ of Concepts\end{tabular} & \textit{\begin{tabular}[c]{@{}l@{}}``\textit{How was the first NFT created?}''\\ ``\textit{What is the outlook for NFTs?}''\end{tabular}} \\ \hline
\multirow{2}{*}{\textbf{Final Cause}} & \multirow{2}{*}{\begin{tabular}[c]{@{}l@{}}``\textit{In the sense of end or `that for} \\ \textit{the sake of which' a thing is done.}''\end{tabular}} & \begin{tabular}[c]{@{}l@{}}Significance\\ of Concepts\end{tabular} & \textit{\begin{tabular}[c]{@{}l@{}}``\textit{What is the importance of NFTs?}''\\ ``\textit{Why do we need NFT as a newly-emerging technology?}''\end{tabular}} \\ \cline{3-4} 
 &  & \begin{tabular}[c]{@{}l@{}}Real-world\\ Consequences\end{tabular} & \textit{\begin{tabular}[c]{@{}l@{}}``\textit{What are the benefits of owning an NFT?}''\\ ``\textit{What is the significance of NFTs in the current capitalist society?}''\end{tabular}} \\ \hline
\end{tabular}%
}
\end{table*}

\subsubsection{Integrating Epistemological Framework}

Here we elucidate the process of generating follow-up questions using the \textit{Four Causes} framework within our system, leveraging the capabilities of ChatGPT. It is worth noting that while a foundational understanding of the Four Causes is not a prerequisite for users, the questions are strategically designed to guide their inquiry towards more complex and contextually rich content.

In our initial evaluation, we assessed ChatGPT's ability to comprehend and apply the Aristotelian Four Causes. 
The results were promising, demonstrating ChatGPT's general adeptness in providing illustrations for each cause in relation to specific concepts or entities. Subsequent to this analysis, we investigated the system's proficiency in crafting follow-up questions anchored in the Four Causes framework. 
ChatGPT exhibited remarkable efficiency in generating inquiries that were not only relevant to the ongoing conversation but also adeptly integrated aspects from the most recent query, thus maintaining a coherent and contextually relevant learning trajectory.

However, it was observed that ChatGPT sometimes struggled to accurately assign the appropriate cause category to each question and to prioritize certain causes over others based on the context or scenario at hand. This suggests a potential area for refinement in the underlying language model, which will be explored more comprehensively in the Discussion section (\S\ref{sec:discussion_diverse}). Despite these challenges, the utilization of the generic model facilitated a structured and in-depth exploration of topics, thereby enhancing the self-learning experience by guiding learners through a more nuanced understanding of the subject matter.

To further refine the guidance provided and enhance its pedagogical value, we incorporated the \textit{persona pattern}, as outlined by White \textit{et al.} \cite{white2023prompt}.
This methodology involves configuring ChatGPT to function as an educational tutor, adapting its way of interacting to more effectively facilitate a learning environment. Within this adapted role, ChatGPT's responsibilities extend beyond merely answering questions; it is also engineered to proactively generate follow-up questions. For each interaction, the directive ``\textit{Provide the top four related follow-up questions based on the previous question using the four causes idea}'' is employed. This strategy ensures the elicitation of a set of questions that are not only relevant but also imbued with educational significance, thus fostering a more engaging and productive self-learning experience.

\subsection{Usage Scenario}

This section illustrates how Emily, a Ph.D. candidate with an interdisciplinary background, efficiently self-learned a new knowledge domain (NFTs) with \name.
In her previous self-study endeavors, Emily had often relied on LLM tools to help her gather and consolidate information to build a foundational understanding of new concepts.
For example, she mentioned using AskYourPDF \footnote{\url{https://askyourpdf.com/zh}} and Poe \footnote{\url{https://poe.com/}} to quickly extract information from references.
Nevertheless, crafting the right questions can be time-consuming, and although some LLM tools offer follow-up questions, they often lack relevance or do not align with her interests. 
As a result, these tools typically only provide hints or keywords, leaving her to rely on traditional search engines or to read original materials for in-depth learning.

Upon adopting \name, Emily's learning process evolved. She first examined the \textit{Concept Graph View} in the upper right corner to gain an overview of the core concepts. 
Then, with the help of the navigation bar, she quickly reviewed the \textit{Embedded Content View} in the upper left corner. 
As a beginner in NFTs, Emily zeroed in on key concepts and sought clarification through the \textit{Q\&A Conversation View}, asking, ``\textit{Can you explain what fungible means in one sentence?}'' 
After reading the reply and realizing that each NFT's price varies--a topic she found intriguing--she inquired further, ``\textit{How do NFT creators determine their products' prices?}'' 
After receiving this response, she engaged in five turns of interaction with the AI chatbot on the topic. 
Under the guidance of the follow-up questions, she conducted an in-depth exploration of NFT pricing and the influencing factors, achieving a comprehensive understanding.

By reviewing the \textit{Tree Map View}, Emily noted that she had explored the factors affecting NFT pricing from various angles, including the personal expectations of artists, the volatility of cryptocurrency prices, and the pricing differences between international and local markets. 
She also understood how to address price volatility.
Satisfied with the depth and breadth of her exploration, Emily then started on a new topic—how NFTs influence society—and pursued deep self-learning through a workflow similar to the one she had previously followed.

\section{Evaluation}

We conducted a within-subject study to evaluate whether \textit{CausaDisco} enhances knowledge sensemaking during self-learning.
Specifically, we aimed to investigate the following three research questions (RQs):

\textbf{RQ1:} How does \textit{CausaDisco} foster interactivity with the LLM chatbot?

\textbf{RQ2:} How does \textit{CausaDisco} facilitate sensemaking in self-directed learning?

\textbf{RQ3:} How do users perceive the usefulness and experience of interacting with \textit{CausaDisco}?

\subsection{Participants}

Our study involved 36 participants (19 male, 17 female) ranging in age from 20 to 36 years old (M = 26.31, SD = 3.01).
We recruited participants through convenience and snowball sampling, using criteria similar to those of our formative study.
Participants had diverse educational backgrounds, with 13 holding bachelor's degrees, 15 holding master's degrees, and 8 holding doctorates.
Most participants (n = 25) specialized in STEM fields, while others came from the humanities and social sciences (n = 5), interdisciplinary studies (n = 3) and medical fields (n = 2).
Participants also had varying levels of academic training experience (more than five years: n=11; three to five years: n=13; fewer than three years: n=12). Familiarity with LLM chatbots ranged from extremely familiar (n = 3) and very familiar (n = 15) to somewhat familiar (n = 11), somewhat unfamiliar (n = 6) and neutral (n = 1).

\subsection{Procedure}

We employed a within-subject experimental design in which participants independently completed two self-learning tasks, each focused on a distinct concept: NFTs \cite{NFTeva} and semiotics \cite{semiotics}.
To minimize order bias, we counterbalanced the order of exposure to our system, \textit{CausaDisco}, and a baseline condition (using participants' preferred tools), resulting in 4 (= 2 x 2) conditions \cite{guo2023grafs, suh2023sensecape}.
Participants were encouraged to engage in active learning through conversational interaction with the LLM chatbot in both conditions  (see Table~\ref{tab: FQ_com} in Appendix~\ref{sec: log}).

The study began with participants providing informed consent and completing a pre-study survey with demographic questions. They then engaged in two self-learning tasks, using either \textit{CausaDisco} or their chosen baseline tools to explore one of the assigned topics. The procedure for self-learning tasks involved four distinct sessions:

\begin{enumerate}
\item 
\textit{Introduction:} A brief overview of the research background and experimental protocols.

\item
\textit{Pre-Task Exercise:} Participants were given 10 minutes for free exploration of the assigned topic.

\item
\textit{Self-Learning Tasks:} Participants engaged in each self-learning task for 20 minutes using either \textit{CausaDisco} or their chosen baseline tools.

\item
\textit{Assessment:} Participants engaged in a 15-minute quiz comprising three multiple-choice and three open-ended questions. This quiz, developed in collaboration with domain experts in NFTs and semiotics, assessed the participants' understanding of the presented concepts.
Following the quiz, participants completed a post-task survey using a seven-point Likert scale to evaluate how each condition supported their sensemaking during self-learning.
The session concluded with semi-structured interviews where participants compared \textit{CausaDisco} and the baseline condition, providing feedback on the system design.
\end{enumerate}

The entire experiment lasted approximately 90 minutes.
The 20-minute duration for each self-learning task was determined based on observations from a pilot study with team members.

\begin{figure*}[tb!]
\centering
  \includegraphics[width=\linewidth]{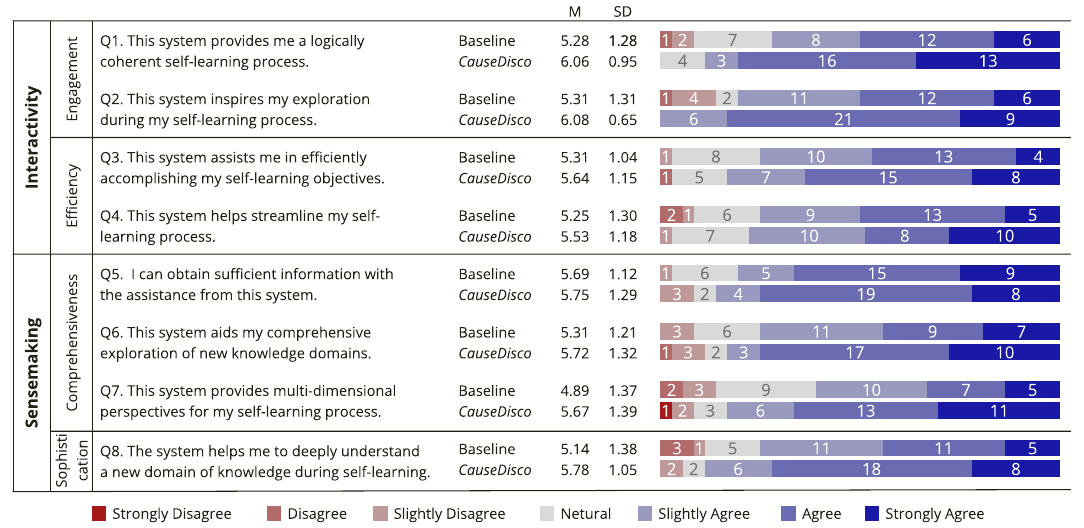}
  \caption{This figure presents participants' subjective ratings (seven-point Likert scale) comparing \textit{CausaDisco} and a baseline on interactivity and sensemaking.} 
  \label{fig: design_eva}
\end{figure*}

\subsection{Measures}

We evaluated participants' perceptions of interactivity, sensemaking, and the usability and design of \textit{CausaDisco}. All measures were rated using a seven-point Likert scale (1 = Strongly Disagree, 7 = Strongly Agree).

\subsubsection{Interactivity Measures}

To assess how \textit{CausaDisco} enhanced interactivity, we analyzed both objective measures and subjective participant perceptions.
We began by analyzing the dialogue records generated during the self-learning tasks, counting and comparing the number of dialogue turns between participants and LLMs in both the \textit{CausaDisco} and baseline conditions.
This analysis provides an objective measure of interaction frequency.

To evaluate participants' subjective experiences of interactivity, we examined two key dimensions: perceived \textit{engagement} and \textit{efficiency} (Fig.~\ref{fig: design_eva}).
\rr{Each dimension comprised two subdimensions. Engagement was assessed by participants' perceived logical coherence of system-generated follow-up questions (\textbf{Q1}) and their perceived inspiration for exploring new knowledge domains (\textbf{Q2}).
Efficiency was evaluated by participants' perceived ability to efficiently accomplish self-learning objectives (\textbf{Q3}) and the perceived streamlining of the learning process (\textbf{Q4}).}

\subsubsection{Sensemaking Measures}

Building upon the sensemaking model proposed by Pirolli \textit{et al.}~\cite{pirolli2005sensemaking}, we assessed how
\textit{CausaDisco} facilitates sensemaking through both subjective and objective measures. For subjective evaluation, we examined two key dimensions: perceived \textit{comprehensiveness} and \textit{sophistication} (Fig.~\ref{fig: design_eva}).
\rr{Comprehensiveness was evaluated by participants' perceived ability to obtain sufficient information with the system (\textbf{Q5}), their perceived support for comprehensive exploration of new knowledge domains (\textbf{Q6}), and their perceived acquisition of multi-dimensional perspectives during self-learning (\textbf{Q7}). Sophistication was assessed by participants' perceived deepening of understanding within a new knowledge domain (\textbf{Q8}).}

For objective evaluation, we recruited domain experts---two Ph.D. students in Fintech and two with master's degrees in communications---to blindly evaluate participants' quiz responses. Each participant's final score was calculated as the average of two expert ratings. The multiple-choice questions assessed the comprehensiveness of understanding, while short-answer questions evaluated the sophistication of knowledge application.

\subsubsection{Usability and Design Measures}

We evaluated how users perceived \textit{CausaDisco}'s usability and design using the technology acceptance model \cite{venkatesh2008technology}.
Participants rated their perceptions of the system's ease of use, ease of learning, and intention to recommend \textit{CausaDisco} on a seven-point Likert scale.
Furthermore, we assessed participants' perceived intuitiveness of the system, encompassing both interface and interaction design elements.

\begin{table*}[tb!]
\caption{Within-subject comparisons of participants' subjective ratings in the two conditions. Measures include perceived engagement, efficiency, comprehensiveness, and sophistication. Statistical significance: * $p < 0.05$, ** $p < 0.01$, *** $p < 0.001$, {\color{gray}*} $p < 0.1$ (marginally significant).}
\label{tab: eva_sta}
\resizebox{\textwidth}{!}{%
\renewcommand{\arraystretch}{1.5}  
\begin{tabular}{l 
                l 
                >{\centering\arraybackslash}m{2cm} 
                >{\centering\arraybackslash}m{2cm} 
                >{\centering\arraybackslash}m{1cm}   
                >{\centering\arraybackslash}m{1cm}   
                >{\centering\arraybackslash}m{1cm}   
                >{\centering\arraybackslash}m{1.2cm}  
                }
\hline
\multirow{2}{*}{\textbf{Measures}} & \multirow{2}{*}{\textbf{Attributes}} & \textbf{Baseline} & \textit{\textbf{CausaDisco}} & \multicolumn{4}{c}{\textbf{Statistics}} \\ \cline{3-8} 
 &  & \multicolumn{1}{c}{\textbf{Mean/SD}} & \multicolumn{1}{c}{\textbf{Mean/SD}} & \multicolumn{1}{c}{\textit{\textbf{t}}} & \multicolumn{1}{c}{\textit{\textbf{df}}} & \multicolumn{1}{c}{\textit{\textbf{p}}} & \multicolumn{1}{c}{\textbf{Sig}} \\ \hline

\multirow{2}{*}{\textbf{Engagement}} & Logically Coherent & 5.28/1.28 & 6.06/0.95 & 3.08 & 35 & 0.001 & ** \\ 
 & Inspiring & 5.31/1.31 & 6.08/0.65 & 3.08 & 35 & 0.001 & ** \\ \hline

\multirow{2}{*}{\textbf{Efficiency}} & Efficiency Enhancement & 5.31/1.04 & 5.64/1.15 & 1.83 & 35 & 0.038 & * \\ 
 & Streamlined Process & 5.25/1.30 & 5.53/1.18 & 1.06 & 35 & 0.149 &  \\ \hline

\multirow{3}{*}{\textbf{Comprehensiveness}} & Sufficient Information & 5.69/1.12 & 5.69/1.31 & 0 & 35 & 0.500 &  \\
 & Comprehensive Exploration & 5.31/1.21 & 5.72/1.32 & 1.46 & 35 & 0.076 & {\color{gray}*} \\  
 & Multi-Dimensional & 4.89/1.37 & 5.67/1.39 & 2.38 & 35 & 0.011 & * \\ \hline

\textbf{Sophistication} & In-Depth Understanding & 5.14/1.38 & 5.78/1.05 & 2.44 & 35 & 0.009 & ** \\ \hline

\end{tabular}%
}
\end{table*}

\section{Results}

This section presents findings from our analysis of participant survey responses, prototype interaction logs, and interview feedback.
To compare quantitative data from surveys and system logs, we used paired samples t-tests between the two conditions, except when Shapiro-Wilk tests indicated deviations from normality.
In those cases, we used the non-parametric Wilcoxon Signed Rank test.

\subsection{Interactivity Enhancement (RQ1)}

To evaluate \textit{CausaDisco}'s interactivity, we assessed both perceived engagement and efficiency using a combination of objective and subjective measures (see Table \ref{tab: eva_sta} and Fig.~\ref{fig: compare}).

\subsubsection{Engagement}

Participants' ratings of logical coherence are significantly higher in the \textit{CausaDisco} condition ($M = 6.06$, $SD = 0.95$) than in the baseline condition ($M = 5.28$, $SD = 1.28$, $p < .01$).
Similarly, participants' ratings indicated significantly higher inspiring of the \textit{CausaDisco} condition ($M = 6.08$, $SD = 0.65$) than the baseline condition ($M = 5.31$, $SD = 1.31$, $p < .01$).
Aligned with the self-rating results, participants engaged in significantly more dialogue turns in the \textit{CausaDisco} ($M = 6.97$, $SD = 2.22$) than in the baseline condition ($M = 4.08$, $SD = 1.86$, $p < .01$).

Qualitative data from participant interviews corroborates these quantitative findings.
Many participants were impressed by \textit{CausaDisco}'s epistemologically informed follow-up questions, describing them as ``\textit{more logically structured, providing a clear train of thought}'' (P35) and ``\textit{more human-like with high-quality}'' (P34).
Furthermore, participants reported that \textit{CausaDisco} facilitated the discovery of new points of interest.
For instance, P25 stated, ``\textit{The follow-up questions and concept graph of CausaDisco are eye-openers! It is like having a curiosity engine--the more I use it, the more I want to keep chatting with the AI.}'' 
The combination of quantitative and qualitative results provides compelling evidence for \textit{CausaDisco}'s effectiveness in enhancing user engagement with LLM chatbots.

\subsubsection{Efficiency}

Participants reported significantly higher efficiency enhancement in the system condition  ($M = 5.64, SD = 1.15$) compared to the baseline condition ($M = 5.31, SD = 1.04, p < .05$). Participants in the system condition, on average, rated the streamlined process 5.53 ($SD = 1.18$), higher than the baseline condition ($M = 5.25, SD = 1.30$), but there are no significant differences.

The results underscore the potential of our system to enhance user efficiency in a significant yet implicit manner.
Although participants reported perceiving an increase in their efficiency, this improvement was not attributed to a more streamlined process.
As P4 remarked, ``\textit{The follow-up questions of CausaDisco are spot-on and boost my self-learning efficiency. I learn faster without having to rack my brain thinking up questions myself. It is a real time-saver.}''
This suggests that our system may boost participants' efficiency by delivering precisely the questions they require at the moment.

However, the underlying logic and mechanism responsible for generating the follow-up questions are not fully comprehensible to participants.
For instance, P29 noted, ``\textit{I often pause to consider the underlying logic of the follow-up questions and the learning trajectory in the tree map. The deeper I go with the questions, the more I find myself taking these little moments to reflect.}''
This uncertainty may lead participants to doubt the effectiveness of the questions in streamlining their self-learning.
This hypothesis is corroborated by the lack of significant findings in streamlined process,  for future research and system improvements to better support users' self-learning.

\begin{figure}[tb!]
\centering
  \includegraphics[width=\linewidth]{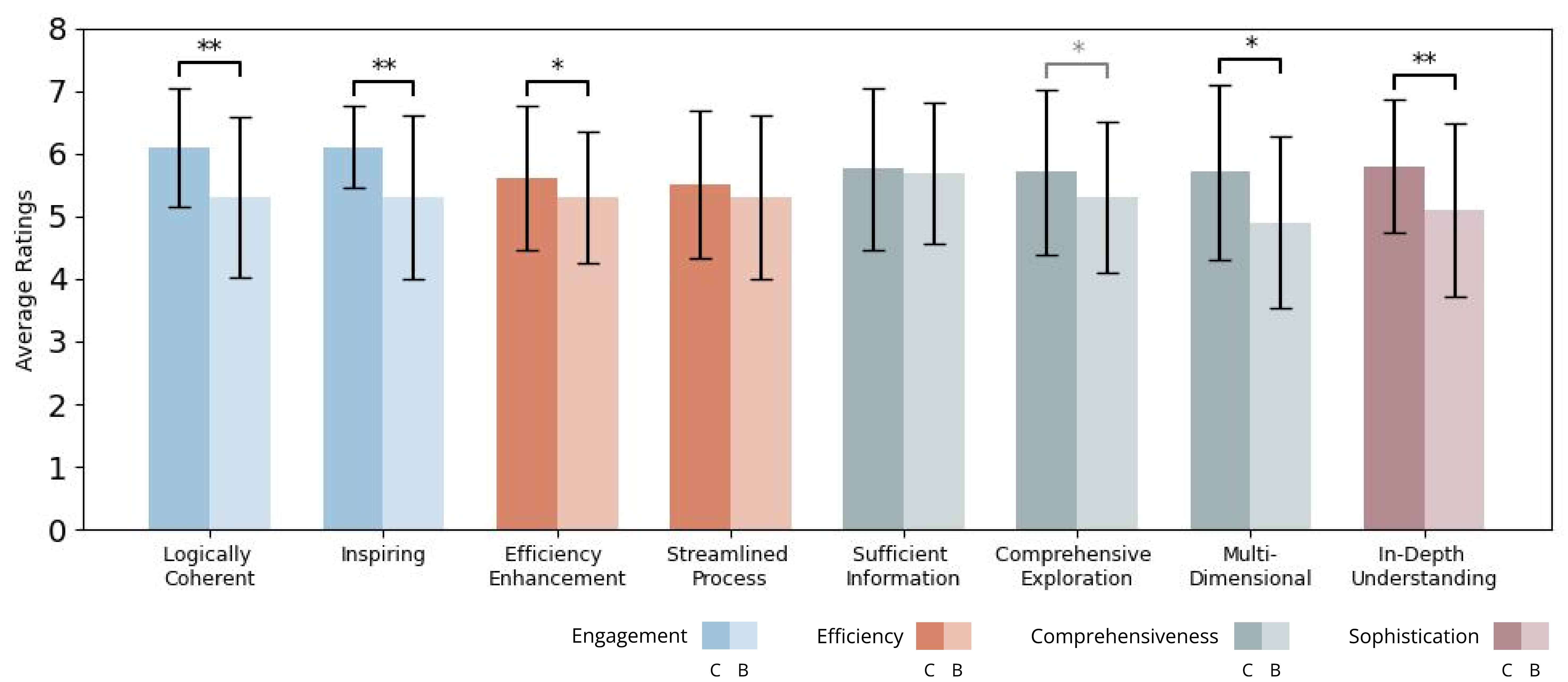}
  \caption{Within-subject comparison of participants' subjective ratings of key measures (engagement, efficiency, comprehensiveness, and sophistication) across two conditions. \rr{Error bars represent between-subjects standard error}; asterisks indicate statistical significance from paired t-tests (*p < 0.05, **p < 0.01). The difference in perceived comprehensive exploration, though not statistically significant, approached significance and is highlighted in gray.} 
  \label{fig: compare}
\end{figure}

\subsection{Sensemaking Support (RQ2)}

To assess how \textit{CausaDisco} facilitates sensemaking, we compared participants' perceived \textit{comprehensiveness} and \textit{sophistication} when processing information under the two conditions (see Table \ref{tab: eva_sta} and Fig.~\ref{fig: compare}).

\subsubsection{Comprehensiveness}
We found no significant differences between conditions regarding sufficient information (the system condition: $M = 5.69, SD = 1.31$; the baseline condition: $M = 5.69, SD = 1.12; p > .05$). However, participants reported marginally significantly higher comprehensive exploration in the system condition ($M = 5.72, SD = 1.32$) compared to the baseline condition ($M = 5.31, SD = 1.21, .05 < p < .10$). They also self-reported higher multi-dimensional ratings in the system condition ($M = 5.67, SD = 1.39$) compared to the baseline condition ($M = 4.89, SD = 1.37, p < .05$). Participants' objective quiz scores in the system condition are, on average, 2.00 ($SD = 0.93$), a little lower than in the baseline condition ($M = 2.25, SD = 0.89$), but there is no significant difference ($p > .05$).
\rr{The results demonstrate that our system did successfully deliver multi-dimensional information to users. However, the scope of the information was constrained by the specific study materials provided. Consequently, participants expressed mixed opinions on whether the new information was sufficient for a fully comprehensive understanding. 
 
In line with our goal of offering diverse perspectives on unfamiliar topics,}
P30 shared, ``\textit{Compared to my usual tools, CausaDisco's follow-up questions focus on key ideas while encouraging broader thinking. Along with the concept map, which gives me a big-picture view of concept connections, CausaDisco helps me grasp new information in a more well-rounded way.}''
This achievement also aligns with our primary objective of providing a comprehensive self-learning environment.

However, we encountered limitations in conclusively proving that our system offers sufficient information.
For instance, P9 commented, ``\textit{It seems all info CausaDisco gives, including the follow-up questions and responses, comes straight from the study materials. That's good for accuracy, but adding some outside facts with sources would help my learning even more.}''
This perceived lack of information likely arises from the contrasting exploration approaches between the two conditions.
\textit{CausaDisco} guides users through a specific article, generating precise, content-based follow-up questions and responses for accuracy.
Conversely, the baseline condition allows unrestricted exploration.
This key difference--\textit{CausaDisco}'s focused, article-centric approach versus the baseline's open-ended exploration--likely created an impression of limited information within our system.

\subsubsection{Sophistication}
Participants reported significantly higher in-depth understanding ratings in the system condition ($M = 5.78$, $SD = 1.05$) compared to the baseline condition ($M = 5.14$, $SD = 1.38$, $p < .01$).
Subjective quiz scores were also higher in the system condition ($M = 5.06$, $SD = 0.98$) than in the baseline condition ($M = 4.56$, $SD = 1.12$, $.05 < p < .1$), although this difference was only marginally significant.

The significant improvement in perceived in-depth understanding, coupled with the trend towards better quiz performance, indicates that \textit{CausaDisco} is likely promoting a more sophisticated exploration during self-learning.
As P1 put it, ``\textit{CausaDisco, especially the follow-up questions and the tree map, lets me really dive deep into a topic.}''
This aligns with our goal of enabling users to seamlessly connect and integrate knowledge through dialogue-driven interaction.

\rr{It is important, however, to acknowledge that subjective perceptions of understanding do not always align with objective outcomes. Future research should further evaluate the system's efficacy in not only deepening conceptual understanding but also supporting the practical application of acquired knowledge.}

\begin{figure*}[tb!]
\centering
  \includegraphics[width=\linewidth]{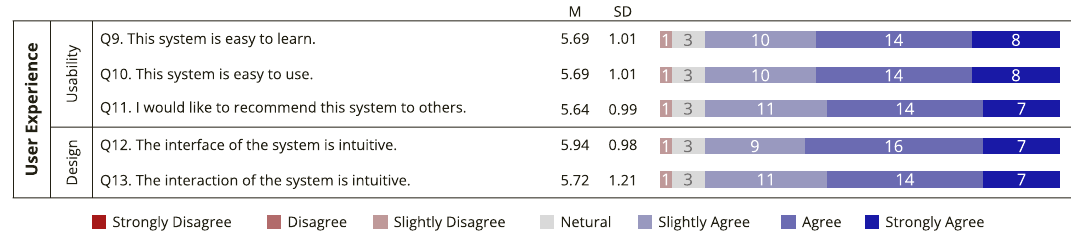}
  \caption{This figure illustrates participants' subjective ratings of usability and design for \textit{CausaDisco}, which were measured using a seven-point Likert scale.} 
  \label{fig: ux_eva}
\end{figure*}

\subsection{Usability and Design (RQ3)}

We also employed a seven-point Likert scale and conducted semi-structured interviews to assess the usability, interface, and interaction design of \textit{CausaDisco} (see Fig. \ref{fig: ux_eva}). 

\subsubsection{Usability}
More than half of the participants strongly agreed that our system is easy to learn and use (\textbf{Q9-10}). 
During the five- to ten-minute introduction, nearly all could understand the main functions of each view of the system and quickly became familiar with the interactive operations through free exploration.
Furthermore, most participants were willing to recommend \textit{CausaDisco} to other self-learners (\textbf{Q11}). 
Only two participants (P23 and P8) were less inclined to do so, mainly because they desired more flexible and personalized interaction. 
S4 explained: ``\textit{I hope the system can further enhance the design and functionality of the tree map by allowing users more freedom to annotate or organize, which could be more helpful}.''

\subsubsection{Design}
Participants generally found the interface design to be intuitive (\textbf{Q12}). 
The \textit{Q\&A Conversation View} and the \textit{Concept Graph View} were particularly well-received. 
For instance, P1 and P4 mentioned, ``\textit{The Concept Graph helps provide me with a big picture and clear direction for active self-learning}.'' 
Additionally, several participants pointed out that the \textit{Tree Map View} helped them trace their learning journey and adjust their query strategies in a timely manner. 
P7 said, ``\textit{The Tree Map View is highly beneficial as it facilitates the efficient organization of the questions I pose, aiding in cognitive training to improve question formulation}.'' 
Furthermore, some participants noted that the \textit{Embedded Content View} alleviated their distrust of the AI chatbot, enabling them to validate original learning materials efficiently. 
As P5 mentioned, ``\textit{The Embedded Content View with navigation bar also assists me in quickly locating relevant sections}.''

In line with the positive feedback received on the interface design, most participants deemed the interaction design of the system (\textbf{Q13}) as both ``\textit{streamlined and essential}.''
Despite this general consensus, two participants, P22 and P24, offered slight critiques. 
P22, in detail, stated, ``\textit{While the system is generally user-friendly, a thorough introduction is still indispensable. 
Lacking such guidance, one might spend considerable time figuring out the system independently}.''

\section{Discussion}

This research aims to integrate epistemological frameworks with LLMs to enhance self-learning. We introduce \textit{CausaDisco}, an interactive system designed to foster comprehensive exploration and holistic understanding through reflective multi-turn dialogue. Our study investigates \rr{epistemological differences across different interaction patterns}. This leads to the development of a robust framework for generating follow-up questions to enhance learners' sensemaking processes.
Our work contributes to the growing body of research on LLM-based educational agents and human-LLM interaction within the HCI community \cite{suh2023sensecape, zheng2024disciplink, hu2024designing, gao2024fine}, aligning with advocates for inclusive, efficient LLM applications in education \cite{shah2022situating, fiannaca2023programming, zamfirescu2023johnny}.
In the following section, we discuss the design implications derived from developing and evaluating \textit{CausaDisco}, its limitations, and potential directions for future work.

\subsection{Inclusive Design for Diverse LLM Interaction Patterns}

\rr{The advent of LLMs has catalyzed a paradigm shift in education, promising customized learning experiences tailored to individual needs and preferences \cite{mogavi2024chatgpt}.
Achieving this, however,} requires ``\textit{dropping an LLM-based agent as a one-size-fits-all solution}'' \cite{shah2022situating} and understanding how the \rr{diverse interaction patterns between learners and LLMs can inform the design of inclusive and adaptive AI tools for self-learning.}

Our qualitative study, \rr{informed by the ICAP framework for classifying learner engagement~\cite{chi2009active, chi2014icap},} revealed four distinct LLM interaction patterns: \textit{proactive}, \textit{validation-seeking}, \textit{content-focused}, and \textit{receptive}.
\rr{These patterns represent different engagement strategies and underlying epistemological schemas that learners may adopt, varying in interaction frequency, knowledge-seeking breadth, and strategic diversity. The choice of strategy appears to be influenced by contextual factors, prior learning experiences, and epistemological beliefs about the nature, source, and justification of knowledge.
We observed that when learners adopted proactive strategies, they demonstrated higher comfort levels with AI-assisted exploration. Conversely, when employing receptive or content-focused strategies--particularly common among those familiar with traditional, input-oriented pedagogy--their LLM engagement was less exploratory. In our LLM-mediated learning context, more passive engagement strategies correlated with lower overall interactivity and diminished quiz performance.
These findings highlight the need for more inclusive LLM-based learning tools that effectively support learners across diverse engagement strategies.}

Designing educational tools that consider the spectrum of \rr{interaction patterns} is essential.
\rr{Learners exhibiting proactive tendencies} may thrive with LLM-based tools that afford expansive application scenarios, diverse interaction modalities, and opportunities for verifying LLM responses.
Conversely, simply providing powerful affordances may prove insufficient for learners \rr{who tend to interact with LLMs in a more restricted manner, particularly those} less familiar or comfortable with AI-driven learning environments.
Extant research suggests that these learners may benefit from carefully designed incentives, structured onboarding experiences, or ``ease-in'' approaches that gently encourage interaction and foster a sense of self-efficacy \cite{zamfirescu2023johnny, fiannaca2023programming}.
Without such considerations, these learners risk being underserved, unable to fully realize the transformative potential of LLMs.

It is critical to underscore that this analysis does not aim to privilege one \rr{interaction pattern} or epistemological orientations over another.
Rather, it highlights the imperative of acknowledging and accommodating the heterogeneity of learners' cognitive processes and prior experiences when designing for LLM-mediated learning environments.
Effective, usable, and inclusive LLM-based tools must be intentionally designed to cater to this diversity, ensuring that all learners, regardless of their prior experience with AI or their preferred learning approaches, can engage with these tools confidently, comfortably, and meaningfully.

\subsection{Design Implications}

\subsubsection{Supporting Knowledge Construction via Mental Models}

The transformative potential of LLMs in education can be significantly enhanced by adopting an epistemological approach to prompt engineering. While current LLM-based educational tools show promise in assisting self-learners, their effectiveness and generalizability are often constrained by domain-specific designs and assumptions about user expertise \cite{sheng2023knowledge, chen2024stugptviz}.
As LLM tools increasingly shape the educational landscape, it is crucial to democratize access and expand their applicability, particularly in providing cognitive support.
To achieve this, developers and designers should consider integrating well-established epistemological frameworks into LLM design.

Our findings, aligning with educational research, highlight the benefits of incorporating systematic guiding mechanisms in LLM interactions \cite{peters2000does, braaten2010personal, krasmann2020logic}.
By encouraging learners to articulate their reasoning, explore diverse perspectives, and integrate new information with existing knowledge, LLM-based educational tools can inspire deeper cognitive engagement.
Grounded in learning sciences and cognitive principles \cite{aksan2009descriptive, mokhtari2014quantitative, baker2020epistemic}, this approach can enhance critical thinking and facilitate robust knowledge construction.

\subsubsection{Fostering Interactivity for Effective LLM Education}

Current LLM-based educational tools, primarily dialogue-driven, face challenges in maintaining user engagement and reducing dropout rates.
\rr{This challenge mirrors a persistent issue in pedagogical research: the difficulty of designing activities that foster constructive and interactive engagement~\cite{chi2018translating}.} 
While the HCI community has explored promising directions through teacher-guided approaches \cite{kumar2023impact} and multi-round dialogues \cite{hu2024designing}, our findings indicate the need for more comprehensive strategies to support LLM-mediated sensemaking in self-learning.

Our research demonstrates that systematic follow-up questions, enhanced by concept graphs, effectively stimulate exploration and promote deeper understanding.
However, to cultivate a truly immersive and engaging self-learning environment, we must consider diverse interaction patterns.
This can be achieved by, for instance, embodying LLMs with distinct personas~\cite{liu2024personaflow}, incorporating interactive visualizations \cite{gao2024fine}, and integrating gamification elements \cite{lacerda2024gamified}.
Such approaches not only cater to diverse \rr{interaction patterns} but also mitigate cognitive load for AI novices.

\subsubsection{Managing Learners' Sensemaking Process}

To create more effective learning experiences, designers and developers could address the distinct stages of the sensemaking process in learning.
This process, which encompasses information retrieval, concept comprehension, knowledge integration, and application, engages both inductive and deductive reasoning \cite{pirolli2005sensemaking, suh2023sensecape, zheng2024disciplink}.

Our research demonstrates that visualizing learners' chains-of-query and promoting metacognition enhance understanding for self-learners.
Based on these findings, we propose developing responsive and adaptive LLM-based educational tools that explicitly support and display various sensemaking stages.
For instance, implementing a meta-layer to track and visualize the learner's granular learning steps can further boost self-awareness and learning efficacy. Such features would enable learners to monitor their progress and adjust their strategies more efficiently.

\subsection{Limitations and Future Work}

Although our study comprehensively investigated participants' LLM interactions during self-learning and identified their epistemological differences to inform the development of \textit{CausaDisco}, it is not without limitations.

\subsubsection{Method Limitations}

As with many qualitative studies, our findings are limited by a modest sample size \cite{hammarberg2016qualitative}, potential researcher biases \cite{galdas2017revisiting}, and the as-yet unconfirmed generalizability of our results \cite{hadi2022users}.
\rr{All of our participants--well-educated adults with substantial academic and self-learning experience--had a clear sense of ``what they should know,'' even though some struggled to ask specific questions.
While our sample provided valuable initial insights, the dynamics of LLM-mediated self-learning likely vary across different educational backgrounds. Learners with less formal education potentially require different forms of support.}
Future work could therefore prioritize validating and extending our taxonomy of self-learners’ interaction patterns and epistemological differences by: (1) including larger, more diverse samples; (2) exploring a wider range of subject domains; and (3) integrating quantitative methods.
\rr{These efforts will inform the development of LLM-based tools that more effectively support learners across different educational backgrounds.}

\subsubsection{Expanding Personalization}

While most participants appreciated the lightweight interactions offered by \textit{CausaDisco}, they also expressed a desire for a more customized experience.
This resonates with Butcher's call~\cite{butcher2011self} for cognitive personalization technologies to support meaningful analysis and coherent understanding among learners with varying levels of prior knowledge.
Our evaluation revealed that participants saw the concept graph not merely as an informative display but as a potentially powerful navigational tool for exploring learning materials.
Thus, they envisioned more flexible interaction features that enable cross-view connections between reading materials, dialogue with the LLM, and the concept graph itself. This user-driven design approach would offer an alternative, logically structured pathway through the material, rather than organizing all keywords according to the author's logic.
Future development will focus on enhancing interactivity, particularly within the concept graph, to create richer ``revisit'' channels \cite{pirolli2005sensemaking} for personalized sensemaking, ultimately improving both comprehension and retention.

\subsubsection{Four Causes Across Diverse Disciplines} \label{sec:discussion_diverse}

In our prototype, we employed ChatGPT to generate follow-up questions without specifying a particular cause from Aristotle's framework. This approach leverages the inherent word associations within the language model to generate contextually appropriate questions, without assessing and understanding the differential significance attributed to each of Aristotle's \textit{Four Causes} within these diverse fields.

Notably, not every discipline inherently involves all four causes. The significance of each cause can vary greatly across different domains, with some causes holding negligible interest or relevance in certain knowledge fields.
For instance, Aristotle, in his \textit{Metaphysics}, posited that lunar eclipses lack a final cause and, strictly speaking, do not pertain to ``matter'' in the conventional sense. Similarly, within undergraduate physics education, the hypothetical final cause of gravity, despite its speculative existence, generally does not engage the curiosity or concern of students. This disinterest in the final cause is so prevalent within scientific discourse that some asserted, ``\textit{Math rejects the final cause.}''
Therefore, to achieve a more targeted and nuanced guidance, a refinement of the language model is imperative. This involves integrating a sophisticated understanding of the varying weights and significances of the \textit{Four Causes} across different disciplines into the model, thereby enhancing the precision and relevance of the generated follow-up questions.

\section{Conclusions}

\rr{This work investigates the design of inclusive LLM-based educational agents that support diverse interaction patterns. Our initial formative study (N=26) examined self-learners' behaviors and mental models during LLM-mediated self-learning, identifying four distinct interaction patterns. Through an epistemological analysis of these interaction patterns, we identified three key challenges in learner-LLM interaction and derived five corresponding design requirements.} To address these challenges, we developed \textit{CausaDisco}, a dialogue-driven LLM-based system designed to facilitate sensemaking and mental model construction when learners engage with complex information. \textit{CausaDisco} integrates a query taxonomy derived from our formative study with the epistemological principles of Aristotle's Four Causes into its prompt engineering process. This approach generates coherent and contextually relevant follow-up questions that promote deeper understanding. A subsequent within-subjects study (N=36) demonstrated that \textit{CausaDisco} fostered more engaging interactions, enhanced perceived comprehension and learning depth, and provided a more intuitive and effective interface. This research advances our understanding of how LLMs can serve as effective educational agents for sensemaking in self-learning and offers valuable design implications for this emerging class of LLM-mediated learning tools.

\section*{Ethical Approval}

The research involving human participants was performed in accordance with the principles of the Declaration of Helsinki. The study protocol was reviewed and approved by the Human and Artefacts Research Ethics Committee (HAREC) at the Hong Kong University of Science and Technology (HKUST), under the reference number HREP-2024-0043. Verbal informed consent was obtained from all participants included in the study.

\bibliographystyle{ACM-Reference-Format}
\bibliography{sample-base}


\appendix
\onecolumn

\section{Formative Study}
\label{apx: formative}

\subsection{Demographics of Semi-structured Interview Participants}

\begin{table*}[htb!]
\caption{The table presents demographic information (age, gender, education) and expertise of formative study participants, along with their years of academic research (YAR), familiarity with large language models (LLM), time spent on self-learning tasks (SLT), quiz time (Quiz T), and quiz scores.}
\label{tab: formative}
\renewcommand{\arraystretch}{1.25}
\resizebox{\textwidth}{!}{%
\begin{tabular}{cccccccccc}
\hline
\textbf{ID} & \textbf{Gender} & \textbf{Age} & \textbf{Edu. Background} & \textbf{Expertise} & \textbf{YAR} & \textbf{LLM Familiarity} & \textbf{SLT (mins)} & \textbf{Quiz T (mins)} & \textbf{Quiz Score} \\ \hline
P1 & Female & 29 & Doctorate degree & Social Science & 5+ & Somewhat Familiar & 23 & 10 & 7/10 \\
P2 & Female & 26 & Bachelor's degree & STEM & 3-5 & Somewhat Familiar & 40 & 11 & 7/10 \\
P3 & Female & 29 & Master's degree & Social Science & $\leq 3$ & Somewhat Familiar & 21 & 9 & 8/10 \\
P4 & Female & 23 & Master's degree & Interdisciplinary Studies & $\leq 3$ & Very Familiar & 20 & 9 & 7.5/10 \\
P5 & Male & 26 & Doctorate degree & STEM & 3-5 & Somewhat Familiar & 18.5 & 7 & 8/10 \\
P6 & Female & 29 & Master's degree & Social Science & $\leq 3$ & Somewhat Familiar & 45 & 15 & 7.5/10 \\
P7 & Male & 29 & Master's degree & STEM & 3-5 & Very Unfamiliar & 18 & 10 & 7/10 \\
P8 & Male & 29 & Master's degree & STEM & N/A & Very Familiar & 18 & 8 & 7/10 \\
P9 & Male & 25 & Master's degree & Interdisciplinary Studies & $\leq 3$ & Very Familiar & 30 & 8 & 9/10 \\
P10 & Male & 25 & Bachelor's degree & STEM & 3-5 & Extremely Familiar & 16 & 4 & 7/10 \\
P11 & Female & 27 & Master's degree & Interdisciplinary Studies & N/A & Somewhat Unfamiliar & 17 & 4 & 7/10 \\
P12 & Male & 32 & Doctorate degree & STEM & 5 + & Very Familiar & 17 & 7 & 8.5/10 \\
P13 & Male & 27 & Doctorate degree & STEM & 3-5 & Very Familiar & 23 & 8 & 5.5/10 \\
P14 & Male & 28 & Doctorate degree & Medical & 3-5 & Somewhat Unfamiliar & 24 & 7 & 5/10 \\
P15 & Female & 27 & Master's degree & Social Science & 3-5 & Somewhat Familiar & 26 & 9 & 5/10 \\
P16 & Female & 26 & Master's degree & STEM & $\leq 3$ & Somewhat Familiar & 20 & 6 & 5.5/10 \\
P17 & Male & 25 & Master's degree & STEM & $\leq 3$ & Very Familiar & 18 & 8 & 8/10 \\
P18 & Female & 29 & Bachelor's degree & Medical & 5 + & Very Unfamiliar & 20 & 15 & 5/10 \\
P19 & Male & 24 & Master's degree & Interdisciplinary Studies & $\leq 3$ & Extremely Familiar & 20 & 6 & 8.5/10 \\
P20 & Female & 27 & Master's degree & Social Science & 3-5 & Somewhat Familiar & 16 & 5 & 8.5/10 \\
P21 & Male & 28 & Master's degree & STEM & 5+ & Somewhat Familiar & 15 & 9 & 5/10 \\
P22 & Female & 24 & Master's degree & STEM & $\leq 3$ & Extremely Familiar & 17.5 & 7 & 6/10 \\
P23 & Female & 32 & Doctorate degree & STEM & 5+ & Somewhat Familiar & 23 & 4 & 9/10 \\
P24 & Female & 28 & Master's degree & Social Science & $\leq 3$ & Somewhat Familiar & 20 & 5 & 6.5/10 \\
P25 & Female & 24 & Bachelor's degree & STEM & $\leq 3$ & Extremely Familiar & 19 & 7 & 5.5/10 \\
P26 & Female & 27 & Master's degree & Social Science & 3-5 & Somewhat Unfamiliar & 42 & 6 & 5/10 \\ \hline
\end{tabular}%
}
\end{table*}

\subsection{Web Page for the Self-learning Task}
\label{apx: probweb}

\begin{figure*}[htb!]
\centering
  \includegraphics[width=\linewidth]{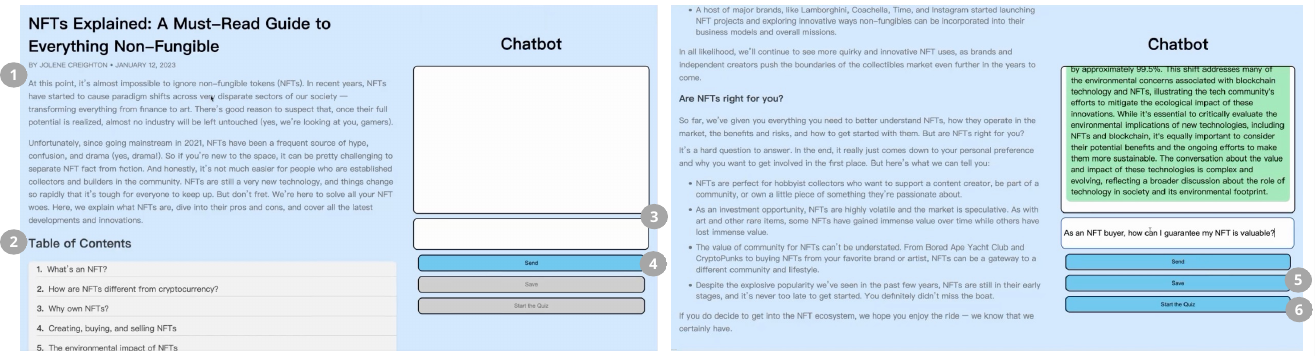}
  \caption{Design of the probe system for self-learning tasks in the formative study. This includes 1) reading materials; 2) a table of contents for navigating different subsections; 3) a question-and-answer chatbot; 4) a \textit{Save} button for downloading dialogue records; and (5) a \textit{Start the Quiz} button to exit self-learning tasks and begin the quiz. The \textit{Save} button and the \textit{Start the Quiz} button become active 15 minutes after entering the web page.} 
  \label{fig: web}
\end{figure*}




\clearpage
\section{Comparing Queries Across Two Conditions}
\label{sec: log}
\vspace{-3mm}

\begin{table}[htb!]
\caption{This table presents follow-up questions on two topics, NFTs and semiotics, generated by \textit{CausaDisco} and GPT-4o (baseline) during user evaluation. Questions in bold represent those selected by users in multi-turn conversations.}
\label{tab: FQ_com}
\renewcommand{\arraystretch}{1.2}
\resizebox{\textwidth}{!}{%
\begin{tabular}{lll}
\hline
\multicolumn{1}{c}{\textbf{Seed Query by Users}} & \multicolumn{1}{c}{\textbf{Condition}} & \textbf{Generated Follow-Up Questions by LLM} \\ \hline
\multirow{14}{*}{\begin{tabular}[c]{@{}l@{}}How do creators of \\ NFTs determine the\\ prices of their products?\end{tabular}} & \multirow{4}{*}{CausaDisco} & \textbf{\begin{tabular}[c]{@{}l@{}}C-1. How do market trends influence the initial pricing strategy of an\\ NFT creator?\end{tabular}} \\
 &  & \begin{tabular}[c]{@{}l@{}}C-2. What impact does the perceived future value have on setting the price of\\ a new NFT?\end{tabular} \\
 &  & \begin{tabular}[c]{@{}l@{}}C-3. In what ways do the creator's reputation and past sales history affect the\\ pricing of their NFTs?\end{tabular} \\
 &  & \begin{tabular}[c]{@{}l@{}}C-4. How does the utility or functionality of an NFT contribute to its valuation\\ by the creator?\end{tabular} \\ \cline{2-3} 
 & \multirow{3}{*}{Baseline} & \begin{tabular}[c]{@{}l@{}}B-1. What are some common platforms for buying and selling NFTs?\end{tabular} \\
 &  & B-2. How do creators market their NFT to attract buyers? \\
 &  & \textbf{B-3. What role does social media play in NFT pricing?} \\ \cline{2-3} 
 & \multirow{4}{*}{CausaDisco} & \textbf{\begin{tabular}[c]{@{}l@{}}C-1-1. How do fluctuations in cryptocurrency value affect the pricing\\ strategies of NFT creators?\end{tabular}} \\
 &  & \begin{tabular}[c]{@{}l@{}}C-1-2. What role do collector and investor behaviors play in shaping market\\ trends for NFTs?\end{tabular} \\
 &  & \begin{tabular}[c]{@{}l@{}}C-1-3. How can NFT creators leverage social media and community\\ engagement to influence market trends and pricing?\end{tabular} \\
 &  & \begin{tabular}[c]{@{}l@{}}C-1-4. In what ways do historical sales data and analytics tools aid NFT\\ creators in understanding and adapting to market trends?\end{tabular} \\ \cline{2-3} 
 & \multirow{3}{*}{Baseline} & \begin{tabular}[c]{@{}l@{}}B-3-1. How do different social media platforms impact NFT pricing differently?\end{tabular} \\
 &  & \textbf{\begin{tabular}[c]{@{}l@{}}B-3-2. What strategies do creators use to leverage social media for pricing\\ their NFTs?\end{tabular}} \\
 &  & \begin{tabular}[c]{@{}l@{}}B-3-3. Can you explain how market trends on social media are identified\\ and analyzed?\end{tabular} \\ \hline
\multirow{14}{*}{\begin{tabular}[c]{@{}l@{}}What is semiotics in \\ concise summary?\end{tabular}} & \multirow{4}{*}{CausaDisco} & \textbf{C-1. What are the key components of a sign in semiotic analysis?} \\
 &  & \begin{tabular}[c]{@{}l@{}}C-2. How does an individual's cultural context shape their understanding\\ of specific signs?\end{tabular} \\
 &  & \begin{tabular}[c]{@{}l@{}}C-2. What role do technological changes play in altering the significance and\\ use of signs in communication?\end{tabular} \\
 &  & \begin{tabular}[c]{@{}l@{}}C-4. How can semiotic knowledge enhance the effectiveness of communication\\ across diverse media formats?\end{tabular} \\ \cline{2-3} 
 & \multirow{3}{*}{Baseline} & B-1. How does semiotics apply to everyday communication? \\
 &  & \textbf{B-2. What are some key theories or figures in semiotics?} \\
 &  & B-3. Can you explain the difference between denotation and connotation? \\ \cline{2-3} 
 & \multirow{4}{*}{CausaDisco} & \textbf{\begin{tabular}[c]{@{}l@{}}C-1-1. How does the relationship between the signifier and the\\ signified influence the interpretation of a sign?\end{tabular}} \\
 &  & \begin{tabular}[c]{@{}l@{}}C-1-2. In what ways can the context alter the perceived relationship\\ between the signifier and the signified?\end{tabular} \\
 &  & \begin{tabular}[c]{@{}l@{}}C-1-3. How do social and cultural factors affect the creation and\\ interpretation of signs?\end{tabular} \\
 &  & \begin{tabular}[c]{@{}l@{}}C-1-4. Can the evolution of language and symbols disrupt the traditional\\ signifier-signified relationship?\end{tabular} \\ \cline{2-3} 
 & \multirow{3}{*}{Baseline} & B-2-1. How do these theories apply to modern digital communication? \\
 &  & \textbf{B-2-2. Can you explain Barthes' concept of myth in more detail?} \\
 &  & B-2-3. What are some practical applications of semiotics in advertising? \\ \hline
\end{tabular}%
}
\end{table}

\end{document}